\input psfig 
\documentstyle[12pt,aaspp4]{article}
\catcode`\@=11 % This allows us to modify PLAIN macros.
\newcommand{\Ref}{\hangindent=20pt \hangafter=1 \noindent}
\newcommand{\StartRef}{\hyphenpenalty=10000 \raggedright}
\def\expa{$^{\rm a)}$}

\def\ergscm{ergs~cm$^{-2}$~s$^{-1}$}
\def\ergs{ergs~s$^{-1}$}
\def\ctss{cts~s$^{-1}$}
\def\nh{cm$^{-2}~$}

\def\Fkev{\footnotesize{Phot cm$^{-2}$ s$^{-1}$ keV$^{-1}$}}
\def\ref{\par\noindent\hangindent 20pt}
\def\sles{\lower2pt\hbox{$\buildrel {\scriptstyle <} 
   \over {\scriptstyle\sim}$}}
\def\sgreat{\lower2pt\hbox{$\buildrel {\scriptstyle >} 
   \over {\scriptstyle\sim}$}}

\def\lsim{\mathrel{\mathpalette\@versim<}}
\def\gsim{\mathrel{\mathpalette\@versim>}}
\def\@versim#1#2{\vcenter{\offinterlineskip
        \ialign{$\m@th#1\hfil##\hfil$\crcr#2\crcr\sim\crcr } }}
\catcode`\@=12 % at signs are no longer letters
\begin{document}
%################### Title and Abstract #######################
\title{Advection-Dominated Accretion Model of the Black Hole
V404$\;$Cyg in Quiescence} 
\author{Ramesh Narayan, Didier Barret, Jeffrey E. McClintock}
\affil{Harvard-Smithsonian Center for Astrophysics, 60 Garden St.,
Cambridge, MA 02138;
rnarayan@cfa.harvard.edu,~barret@cesr.cnes.fr,~jmcclintock@cfa.harvard.edu}

\begin{abstract}

We have analyzed archival ASCA data on the soft X-ray transient source
V404 Cyg in quiescence.  We find that in the energy range 0.7 to 8.5
keV the spectrum is a hard power-law with a photon spectral index
between 1.8 and 2.6 (90\% confidence limits).  We present a model of
V404 Cyg in which the accretion flow has two components: (1) an outer
thin disk with a small annular extent, and (2) a large interior region
where the flow is advection-dominated.  Nearly all the radiation in
the infrared, optical, UV and X-ray bands is from the
advection-dominated zone; the thin disk radiates primarily in the
infrared where it contributes about ten percent of the observed flux.
The spectrum we calculate with this model is in excellent agreement
with the ASCA X-ray data presented here, as well as with previous
optical data.  Moreover, the fit is very insensitive to the choice of
parameters such as black hole mass, orbital inclination, viscosity
coefficient $\alpha$, and magnetic field strength.  We consider the
success of the model to be strong support for the advection-dominated
accretion paradigm, and further evidence of the black hole nature of
V404 Cyg.  We discuss strategies whereby systems with
advection-dominated accretion could be used to prove the reality of
event horizons in black holes.

\end{abstract}

Subject headings: accretion: accretion disks --- binaries: close
--- black hole physics --- X-ray binaries

%################### Introduction #######################
\section{Introduction.}

Soft X-ray transients (SXTs) are among the best-known examples of
accreting black holes (van Paradijs \& McClintock 1995, Tanaka 
\& Shibazaki 1996).  Several of these X-ray binaries have measured 
mass functions larger than $3M_\odot$, which exceeds the maximum mass
of a neutron star, thereby confirming the black hole nature of the
systems.  Since SXTs are among the most accessible black holes for
observational study, much effort has gone into observing and modeling
them.

SXTs occasionally have outbursts lasting several months in which they
flare up to a luminosity comparable to the Eddington limit.  Most of
the time, however, these sources are in a quiescent state with a
luminosity several orders of magnitude below Eddington.

Observations of quiescent systems are difficult, but have nevertheless
been carried out recently for a few systems in the optical, UV and
X-ray bands.  These observations have revealed a serious problem for
models of quiescent SXTs.  Specifically, in at least one system,
namely A0620--00 (McClintock, Horne \& Remillard 1995), the optical and
X-ray data are inconsistent with any accretion model based purely on a
thin accretion disk.  Narayan, McClintock \& Yi (1996, hereafter NMY)
showed that the optical data in A0620--00 can be fitted with a thin
disk only if the inner edge of the disk is at several thousand
Schwarzschild radii; however, such a model cannot explain the X-ray
flux.  On the other hand, if one tries to fit the X-ray data with a
thin disk extending down to the black hole, the spectral shape does
not fit the observations, and the model disagrees strongly with the
optical, UV and EUV constraints.

The outbursts of SXTs are usually explained by means of a thermal
limit cycle model (Mineshige \& Wheeler 1989).  In this model, the
entire accretion flow, from the outer edge down to the last stable
orbit at three Schwarzschild radii, is assumed to be in the form of a
thin disk, both in quiescence and in outburst.  The fact that the
observed spectrum in quiescence is inconsistent with a pure thin disk
is clearly a problem for this model.  Furthermore, the estimated
recurrence time between outbursts disagrees seriously with
observational constraints (Lasota, Narayan \& Yi 1996b).

These difficulties were overcome in a new model of quiescent SXTs
proposed by NMY.  In this model, the accretion flow occurs as a thin
disk only outside a transition radius $r_{\rm tr}$ (we use $r$ to
represent the dimensionless radius in Schwarzschild units), while for
the energetically important region $r<r_{\rm tr}$ the accretion is via
an advection-dominated accretion flow (ADAF).  This model is able to
explain the observed spectrum of A0620--00 (NMY), and at the same time
appears also to predict a consistent recurrence time between outbursts
(Lasota et al. 1996b).  In the model of A0620--00 presented by NMY,
$r_{\rm tr}\sim ~{\rm few}\times10^3$.  However, in the models of V404
Cyg presented in this paper, $r_{\rm tr}>10^4$.  These models are thus
dominated by the ADAF, and the thin disk plays only a minor role.

ADAFs are accretion flows in which most of the energy released via
viscous dissipation remains in the accreting gas rather than being
radiated away.  They get their name from the fact that the bulk of the
energy is advected with the flow.  This is in contrast to a thin
accretion disk, where essentially all the dissipated energy is
radiated locally.  ADAFs can occur in two distinct regimes.

At sufficiently high mass accretion rates $\dot M$, radiation is
trapped by the accreting gas and is dragged into the central star
(Katz 1977, Begelman 1978).  In a comprehensive analysis, Abramowicz
et al. (1988) showed that this regime of ADAFs corresponds to a
fundamentally new branch of accretion.  The branch has, however, not
yet found application to real sources.

At low $\dot M$, a second branch of ADAFs is possible as the viscously
heated gas becomes extremely optically thin and is unable to cool
within the accretion time scale of the flow.  The critical bottleneck
to the cooling is the fact that the gas becomes a two-temperature
plasma (Shapiro, Lightman \& Eardley 1976) so that energy transfer
from ions, which acquire most of the viscously dissipated energy, to
electrons is inefficient.  Most of the energy then remains in the gas
as thermal energy of the ions and is advected into the central star
rather than being radiated.  This regime of ADAFs was briefly
discussed by Rees et al. (1982) when they considered an ``ion torus''
model in which they merged the two-temperature accretion model of
Shapiro et al. (1976) with some new ideas on accretion tori and
funnels (Fishbone \& Moncrief 1976, Abramowicz, Jaroszy\'nski \&
Sikora 1978).  In this model the two-temperature gas
%, which in principle need not accrete according to the authors, 
forms a torus with open funnels through which jets are formed via the
Blandford \& Znajek (1977) mechanism.

Recent theoretical work by a number of authors (Narayan \& Yi 1994,
1995ab, Abramowicz et al. 1995, Chen 1995, Chen et al. 1995, Narayan,
Kato \& Honma 1997, Chen, Abramowicz \& Lasota 1997, see Narayan 1996a, 1997
for reviews) has led to a clear recognition of the low $\dot M$,
optically thin regime of ADAFs as a new and distinct branch of
accretion with dynamically consistent solutions.  A number of the
basic dynamical properties of optically thin ADAFs have been worked
out, and it has been shown that these flows produce X-ray spectra
similar to those observed in many accreting black holes.  Furthermore,
it is now clear that the low $\dot M$, optically thin ADAF branch is
quite distinct from the Shapiro et al. (1976) solution.  The
distinction is important since the Shapiro et al. flow is known to be
violently unstable (Piran 1978).  The new ADAF branch, in contrast, is
either fully stable (Abramowicz et al. 1995, Narayan \& Yi 1995b) or
at most has a weak instability which is no threat to the global
viability of the flow (Kato, Abramowicz \& Chen 1996).

Optically thin ADAFs appear to be perfect for explaining a variety of
observations of accreting black holes, and there have indeed been
several applications to real systems (Narayan, Yi \& Mahadevan 1995,
NMY, Lasota et al. 1996a, Fabian \& Rees 1995, Mahadevan 1996).  In
particular, the NMY model of quiescent SXTs makes use of this solution
to model the region of the accretion flow inside $r=r_{\rm tr}$.  The
present paper describes further work on the NMY model, and is driven
by three recent developments.

First, we have improved the modeling techniques significantly, both in
the description of gas dynamics and in the treatment of radiation
processes.  Whereas the NMY work was based on a self-similar solution
of the gas flow (Spruit et al. 1987, Narayan \& Yi 1994), we now make
use of self-consistent global solutions with proper boundary
conditions (Narayan et al. 1997, Chen et al. 1997).  We have also
improved the treatment of Comptonization.  The details are described
in \S2 below.

Second, we focus on a prediction made by NMY, namely that the SXT
source V404 Cyg in quiescence should have quite a hard X-ray spectrum,
with a photon index of the order of 2.  At the time of that paper, the
data were not good enough to check the prediction.  There was an
indication based on ROSAT data that the spectrum may be extremely soft
with $\alpha\sim7$ (Wagner et al. 1994).  However, a reanalysis of the
same data by NMY gave $\alpha= 4.0_{-1.5}^{+1.9}$.  NMY suggested that
an observation with the ASCA satellite would provide a definitive test
of the model.  V404 Cyg has now been observed by ASCA and the data are
of sufficient quality for this test.  We describe in \S\S3.1, 3.2 our
analysis of the ASCA X-ray data and observations in other wavelength
bands, and show in \S3.3 that the ADAF model is in very good agreement
with the observations.  V404 Cyg is the most robust black hole
candidate among the SXTs, since it has the largest mass function.  It
also has the strongest X-ray flux in quiescence.  For both reasons, it
is the best SXT to pursue for detailed testing of the ADAF paradigm.
For completeness, we present in \S3.4 a reanalysis of A0620--00 using
the improved techniques of this paper.

Finally, Wheeler (1996) has recently argued that the original NMY
model is inconsistent since the $\dot M$ of those models corresponds
to a thermal instability in the outer thin disk.  We show in \S4 that
the new models presented in this paper do not suffer from the
instability and are fully consistent.  We also discuss in this section
future opportunities for further testing the ADAF paradigm, and for
establishing the reality of the event horizon in black holes.

%###################### Modeling Techniques ######################
\section{Modeling Techniques}

\subsection{The Model and its Parameters}

We consider a black hole of mass $M$ accreting gas at a steady mass
accretion rate $\dot M$.  The outer edge of the accretion flow is at a
radius $r=r_{\rm out}$, where $r$, the radius in Schwarzschild units, is
related to the physical radius $R$ by
$$
r={R\over R_S},\qquad R_S={2GM\over c^2}. \eqno (2.1.1)
$$
The angular momentum vector of the binary is inclined at an angle $i$
to the line-of-sight.  For many astrophysical sources we have direct
estimates of $M$, $r_{\rm out}$ and $i$ from the observations.  However,
$\dot M$ is generally not known and needs to be adjusted so as to fit
the observed spectrum.

Following NMY, we model the accretion flow by means of two zones
separated at a transition radius $r_{\rm tr}$.  For $r<r_{\rm tr}$, we
assume that the gas is in the form of an optically thin
two-temperature ADAF.  For $r>r_{\rm tr}$, the accretion occurs
primarily in the form of a thin accretion disk, but with part of the
mass accreting via a corona above the disk.  In our picture, the
accretion stream from the companion star arrives at the disk in a more
or less cool state and initially forms a standard thin disk.  As the
accretion proceeds inward, gas evaporates continuously from the
surface of the disk into the corona, so that with decreasing radius
more and more of the gas shifts from the disk to the corona.  By
$r=r_{\rm tr}$, the entire accretion flow is transformed completely
into the corona, and the coronal flow continues inward as a pure ADAF.
As far as the physics of the flow is concerned, we do not make any
distinction between the ADAF and the corona.  We view the corona
merely as another component of the ADAF, except that it happens to be
at the same radius as an equatorial thin disk.

The evaporation from thin disk to corona probably occurs by a process
similar to that described by Meyer \& Meyer-Hofmeister (1994) or Honma
(1995).  However, there are uncertainties still in the precise
mechanism and it is not possible to estimate from first principles
either the rate of evaporation or the precise location of the
transition radius.  We therefore model the evaporation in an empirical
way and assume that the mass accretion rate in the corona ($\dot M_c$)
and the disk ($\dot M_d$) vary with radius $r$ as
$$
\dot M_{c}=\dot M\left({r_{\rm tr}\over r}\right),
\qquad \dot M_{d}=\dot M-\dot M_{c},\qquad r>r_{\rm tr}.\eqno (2.1.2)
$$
The functional form we have assumed for $\dot M_c$ is fairly
arbitrary, but it is probably not unreasonable.  In any case, the
results presented in this paper are completely insensitive to the
specific choice made in equation (2.1.2), as we show in \S3.3.

We estimate the transition radius $r_{\rm tr}$ on the basis of the
largest velocity $v_{\rm max}$ seen in the H$\alpha$ emission line
from the thin accretion disk (cf. NMY):
$$
r_{\rm tr}={1\over2}\left({c\sin i\over v_{\rm max}}\right)^2.\eqno (2.1.3)
$$

This completes the description of the binary system and the geometry
of the flow.  In addition, we need three parameters to describe the
properties of the accreting gas in the ADAF and the corona:

\noindent
1. We model viscosity by means of the standard $\alpha$ prescription
(cf. Frank, King \& Raine 1992) and write the kinematic viscosity
coefficient as
$$
\nu=\alpha{c_s^2\over\Omega_K}, \eqno (2.1.4)
$$
where $c_s$ is the isothermal sound speed ($p/\rho$) and $\Omega_K$ is
the Keplerian angular velocity.  We assume that the parameter $\alpha$
is independent of $r$.  In most of our models, we set $\alpha=0.3$.

\noindent
2. We assume that the accreting gas in the ADAF and corona consists of
a mixture of particles and magnetic fields in rough equipartition.
Thus, we write the pressure due to the two components as
$$
p_{\rm gas}=\beta p_{\rm tot},\qquad p_{\rm mag}=(1-\beta)p_{\rm tot}, 
\eqno (2.1.5)
$$
with $\beta$ taken to be independent of $r$.  We do not consider
radiation pressure, which is easily shown to be negligible in these
flows (Narayan \& Yi 1995b).  In our calculations, we generally set
$\beta=0.5$, but we also show one model with $\beta=0.95$.  In
principle, if the macroscopic viscosity is entirely due to magnetic
fields, we may expect $\alpha$ and $\beta$ to be related to each
other.  For instance, according to Hawley (1996),
$$
\alpha\sim{1\over2}{p_{\rm mag}\over p_{\rm tot}}={1-\beta\over2}. 
\eqno (2.1.6)
$$
Our standard parameter set, $\alpha=0.3$, $\beta=0.5$, is compatible
with this scaling.

\noindent
3. We include a third parameter $\delta$, which describes the fraction of
the viscous energy which goes directly into electrons.  Various
heuristic arguments (e.g. Shapiro et al. 1976, Phinney 1981, Rees et
al. 1982) suggest that viscous dissipation primarily heats up ions and
that only a fraction $\sim(m_e/m_i)$ of the energy goes directly into
electrons.  With this in mind, we normally set $\delta=10^{-3}$.
However, we also try larger values of $\delta$ in order to test the
sensitivity of the results to this parameter.

\subsection{Dynamics and Energy Balance of the ADAF and Corona}

We employ a logarithmically spaced radial grid to describe the
properties of the ADAF, the corona, and the thin disk.  The cell edges
are at radii $\{l_j\}$, where $l_1=1$ (black hole event horizon) and
$l_n=r_{\rm out}$ (outer edge).  The mid-points of the cells are
denoted by $\{r_j\}$, with $\log r_j=(\log l_{j-1}+\log l_j)/2$.
Cells are identified by the index $j$ of the mid-point, which in this
scheme ranges from 2 to $n$.  Typically, we have ten cells per decade
of radius.  We do not make a distinction between the ADAF and the
corona, but treat them as a single entity.  The only exception to this
statement is that we allow $\dot M$ to change with $r$ in the corona
(cf. eq. 2.1.2), whereas it is independent of $r$ in the pure ADAF
zone.

We model the ADAF as a set of nested spherical shells.  In order to
allow for the flattening of the density profile, we truncate each
shell near the pole in the manner discussed in Appendix A.  Thus, the
shell at $r_j$ extends from polar angle $\theta_{H,j}$ to
$\pi-\theta_{H,j}$, where $\theta_{H,j}$ is calculated from the local
sound speed and angular velocity of the gas (Appendix A).  For
simplicity, we assume that the gas density is constant within each
radial shell.

We take the radial velocity $v(r)$, angular velocity $\Omega(r)$ and
isothermal sound speed $c_s(r)$ from the numerical global solutions
calculated by Narayan et al. (1997) (see also Chen et al. 1997).  Each
model satisfies the basic conservation laws of mass, radial momentum,
angular momentum and energy, along with consistent boundary conditions
at the inner and outer edges of the flow.  On the outside, the
solutions match on to a self-similar form (Narayan \& Yi 1994), while
on the inside the gas makes a sonic transition and falls
supersonically into the black hole, with a zero-torque condition at
the horizon (see Narayan et al. 1997 for details).  The models assume
a pseudo-Newtonian form of the gravitational potential which mimics
some of the properties of a Schwarzschild black hole (Paczy\'nski \&
Wiita 1980).

Each global model is uniquely specified by the value of the viscosity
parameter $\alpha$, the ratio of specific heats $\gamma$ of the
accreting gas, and a function $f(r)$ which describes the nature of
the energy advection.  For a plasma with a constant $\beta$ and an
isotropically tangled magnetic field, Esin (1996) has shown that the
effective $\gamma$ is given by
$$
\gamma={8-3\beta\over6-3\beta}. \eqno (2.2.1)
$$
(Note that this is different from the value quoted in Narayan \& Yi
1995b.)  The function $f(r)$ describes at each radius $r$ the fraction
of the viscously dissipated energy which is advected radially with the
gas.  For the quiescent SXTs which we consider in this paper, the
radiative efficiency is extremely low and so $f(r)$ is practically
equal to unity at all radii.  We set $f(r)=1$ in calculating the
global models, an excellent approximation for the calculations
presented here.

The global solution specifies the values of $v_j$, $\Omega_j$,
$(d\Omega/dr)_j$, and $c_{s,j}$ at the mid-points of the cells of our
model.  Mass conservation directly gives the density:
$$
\rho_j={\dot M(r)\over 4\pi r_j^2v_j\cos\theta_{H,j}}. \eqno (2.2.2)
$$
Note that $\dot M(r)$ is equal to the full $\dot M$ for $r_j<r_{\rm tr}$,
but is equal to the coronal $\dot M_c(r_j)$ for $r_j\geq r_{\rm tr}$ (cf.
eq. 2.1.2).  The viscous heating rate per unit volume is given by 
$$
q_j^+=\rho_j\nu_j\left(r{d\Omega\over dr}\right)_j^2, \qquad
\nu_j=\alpha{c_{s,j}^2\over\Omega_{K,j}}, \eqno (2.2.3) 
$$ 
where $\nu_j$ is the kinematic viscosity coefficient in the $j$th
cell (cf.  eq. 2.1.4).  The heating over the entire shell is then 
$$
Q_j^+=q_j^+V_j,\qquad V_j={4\pi\over3}
(l_j^3-l_{j-1}^3)\cos\theta_{H,j}. \eqno (2.2.4) 
$$

Since we model the accreting gas as a two-temperature plasma
(Shapiro et al. 1976), we need to determine at each radius the ion
temperature $T_i$ and electron temperature $T_e$.  By the
definition of $\beta$, we immediately have one constraint on the
temperatures:
$$
\beta\rho c_s^2=p_{\rm gas}={\rho kT_i\over\mu_im_u}+{\rho kT_e\over\mu_em_u}, 
\eqno (2.2.5)
$$
where $\mu_i=1.23$ and $\mu_e=1.14$ (Narayan \& Yi 1995b).  A second
relation is obtained via the energy balance of the electrons:
$$
\delta Q_j^++Q_j^{\rm ie}=Q_j^-.\eqno (2.2.6)
$$
All quantities in this equation refer to shell $j$ and are integrated
over the shell.  The first term on the left is the direct viscous
heating of electrons (recall the definition of $\delta$ in \S2.1).
The second term is the heating of electrons via Coulomb collisions
with the ions.  Stepney \& Guilbert (1983) have given a convenient
expression for this heating rate as a function of the densities and
temperatures of the electrons and ions.  The term on the right is the
radiative cooling of the gas in the shell, which is computed by the
methods described in \S2.3.  By simultaneously solving equations
(2.2.5) and (2.2.6) in each radial shell we determine $T_{i,j}\equiv
T_i(r_j)$ and $T_{e,j}\equiv T_e(r_j)$ in the various shells.  The
solution is obtained iteratively.

\subsection{Radiative Transfer, Cooling, and Spectrum}

The bulk of the effort goes into calculating the radiative properties
of the ADAF and corona.  Because of the non-local nature of the
radiative interactions (especially Compton scattering), we use the
iterative scattering method (Poutanen \& Svensson 1996, Sunyaev \&
Titarchuk 1985), which is related to the $\Lambda$ iteration method
(Mihalas 1978).  This method allows us to build up the local radiation
field at each point in the flow iteratively and at the same time to
calculate the rate of cooling ($Q_j^-$) of the accreting gas.  As a
by-product the calculation also gives the spectrum seen by an observer
at infinity.

A few approximations are made in the version of the calculations
described below.  This is a non-relativistic code, as dictated by the
fact that we use a pseudo-Newtonian potential (Paczy\'nski \& Wiita
1980) to calculate the global flow solution.  As part of this
approximation, we neglect gravitational redshift, Doppler boosts, and
ray deflections.  For advection-dominated flows the first two effects
tend largely to cancel each other, and one should either include both
self-consistently or not include either.  We have chosen the latter
alternative.  Abramowicz et al. (1996) have recently computed global
solutions in Kerr geometry and it should soon be possible to develop
relativistically consistent radiative models.  Another simplification
is that we set the radial velocity and density to be independent of
$\theta$ (the constancy of $v$ is implicit in eq. 2.2.2).

\subsubsection{Scattering Probability Matrices}

Compton scattering is an important, often dominant, process in
ADAFs.  Since the accreting gas is optically thin, scattering is
extremely non-local, and photons emitted at one radius are capable of
being scattered at practically any other radius.  To deal with this
we calculate the following three matrices which describe the
properties of the scattering:

\noindent
$P_{jk}^{\rm aa}$: The probability that a photon emitted by the ADAF (or
corona, recall we make no distinction) in shell $j$ is scattered by an
ADAF electron in shell $k$.

\noindent
$P_{jk}^{\rm ad}$: The probability that a photon emitted by the ADAF in
shell $j$ irradiates ring $k$ of the thin disk.  We assume that the
irradiating flux is completely absorbed by the disk.

\noindent
$P_{jk}^{\rm da}$: The probability that a photon emitted by ring $j$ of
the thin disk is scattered in shell $k$ of the ADAF.

In principle, each of these matrices is a function of the frequency
$\nu$ of the photon since the scattering cross-section declines with
increasing photon energy in the Klein-Nishina regime.  We ignore the
frequency dependence in calculating the $P_{jk}$, using the
non-relativistic Thomson cross-section $\sigma_T$ for the
cross-section per electron.  The correct cross-section is, however,
automatically included when doing the actual Comptonization
calculation (see \S2.3.4 below).  The error due to this approximation
is therefore minor.

We assume that the radiation emitted at each point in the flow is
locally isotropic.  This assumption is valid for the synchrotron and
bremsstrahlung radiation (except for the effect of Doppler beaming
which is ignored in this treatment), but is a simplification for
Compton scattering.  The reason is that the incident radiation field
at each point of the flow is in general anisotropic, and so the
scattered radiation too would have some residual anisotropy (though
less than in the incident field).  The anisotropy is easily included
in the calculations by generalizing the $P_{jk}$ to $P_{jk,lm}$ where
$l$ and $m$ refer to the orientations of the pre- and post-scattered
photon.  We have avoided this generalization in the interests of
computational speed.

The matrix elements $P_{jk}^{\rm aa}$ and $P_{jk}^{\rm ad}$ are computed by
shooting a large number of rays out of each shell $j$ and following
their scattering histories.  The points of origin of the rays are
distributed uniformly in $\cos\theta$ within each shell; typically we
choose $n_\theta=6$ different values of $\cos\theta$.  At each point
of origin, we select a set of ray directions $\{\chi_l\}$ where
$\chi$ is the angle between the ray direction and the negative of the
local radius vector.  Because the ADAF extends over many decades of
radius, the set of $\{\chi_l\}$ has to be selected with care so that
interactions between all pairs of shells are included.  The specific
choice we make is the following set of backward oriented rays: 
$$
\sin\chi_l=r_{l+1}/r_n,\qquad l=1,2,\cdots,n-2.\eqno (2.3.1) 
$$ 
This set is augmented with a few additional rays in sideways and
forward directions to fill up the entire $4\pi$ solid angle.
Typically, we end up with a set of about 50 values of $\chi_l$.  The
weights of the rays are chosen so as to reflect the solid angle
covered by each.  Finally, for each point of origin and each
$\chi_l$, we choose $n_\phi$ different rays corresponding to
different azimuthal angles relative to the local radius vector;
usually $n_\phi=6$.  Thus, in total, we have over $10^3$ rays per
radial shell.

We calculate the trajectory of each ray and compute the distances
traversed by the ray in various shells (including the shell of
origin).  Since we know the electron densities in the shells (from
$\rho_j$, eq. 2.2.2), we can compute the differential Thomson
cross-sections and can thereby calculate the fraction of the energy in
the ray which is lost by scattering in each shell.  These quantities
are accumulated in $P_{jk}^{\rm aa}$ with the weight appropriate to
the ray.  If a ray intersects the thin disk at any point, it is
assumed that all the residual energy in the ray is absorbed by the
particular local ring segment $k$ of the disk.  In this case, the
matrix element $P_{jk}^{\rm ad}$ is updated.  Note that we do not
include electron-positron pairs in the model.  At the low mass
accretion rates considered in this paper pairs are quite negligible
(cf. Bjornsson et al. 1996, Kusunose \& Mineshige 1996).

In an exactly analogous fashion, we shoot rays off the surfaces of the
various rings $j$ in the thin disk and follow their histories to
calculate the matrix elements $P_{jk}^{\rm da}$.

As an independent calculation, we shoot rays from the ADAF/corona and
the thin disk at the specific inclination angle $i$ of the system and
calculate the following quantities:

\noindent
$E_j^{\rm a}$: The probability that a photon emitted in shell $j$ of the
ADAF or corona at an angle $i$ to the spin axis escapes unscattered
to infinity.

\noindent
$E_j^{\rm d}$: The probability that a photon emitted in ring $j$ of the
thin disk at an angle $i$ to the spin axis escapes unscattered to
infinity.

\subsubsection{Bremsstrahlung Emission}

The net bremsstrahlung cooling per unit volume, $q_{\rm brem}^-$, of a
thermal plasma has been calculated by Stepney \& Guilbert (1983) and
Svensson (1982) (with some slight modifications introduced by Narayan
\& Yi 1995b).  The luminosity from shell $j$ is thus given by
$$
L_{{\rm brem},j}(\nu)d\nu=q_{{\rm brem},j}^-V_j\left({h\over kT_{e,j}}\right)
\exp\left(-{h\nu\over kT_{e,j}}\right)\bar g_{\rm ff}(\nu)d\nu, \eqno (2.3.2)
$$
where $\bar g_{\rm ff}(\nu)$ is a velocity averaged Gaunt factor, which
we take from Novikov \& Thorne (1973).  We normalize $\bar g_{\rm ff}$ so
that the integral of $L_{{\rm brem},j}d\nu$ is equal to $q_{{\rm brem},j}V_j$,
the net cooling of the shell due to bremsstrahlung emission.

Free-free absorption is extremely small in all our ADAF models and is
neglected.

Note: we consider below a number of spectral densities similar to
$L_{{\rm brem},j}(\nu)$.  All of these are functions of $\nu$ and have units
of ergs s$^{-1}$ Hz$^{-1}$.  To simplify the notation we will omit the
argument $(\nu)$ unless it is essential.

\subsubsection{Synchrotron Emission}

The electrons in ADAF models are typically quasi-relativistic
($kT_e\sim m_ec^2$) and therefore emit cyclo-synchrotron radiation in
the local magnetic field.  The cyclo-synchrotron emissivity
$j_{\rm synch}(\nu)$ of a thermal plasma has been worked out by a number
of authors in various limits (Pacholczyk 1970, Petrosian 1981,
Takahara \& Tsuruta 1982, Mahadevan, Narayan \& Yi 1996), and
convenient fitting formulae for the general case are presented by
Mahadevan et al. (1996).  The emission from shell $j$ in the optically
thin limit is thus easily calculated: $j_{{\rm synch},j}(\nu)V_j$.  The main
complication is that the synchrotron emission is self-absorbed.  We
allow for this as follows.

We step outwards, starting with the innermost shell, and calculate
for each shell $j$ the net outgoing synchrotron radiation
$L_{{\rm synch},j}$ due to all shells interior to and including shell $j$.
To understand how we calculate $L_{{\rm synch},j}$, consider the
changes that occur when we step from shell $j$ to shell $j+1$:

\noindent
(i) A fraction $P_{j,j+1}^{\rm aa}$ of the outgoing $L_{{\rm synch},j}$ is
scattered in shell $j+1$.

\noindent
(ii) The shell $j+1$ emits synchrotron radiation with a net
luminosity of $j_{{\rm synch},j+1}V_{j+1}$.  This is added to the previous
luminosity.

\noindent
(iii) Some part of the radiation is absorbed by the gas.  We handle
the absorption in an approximate fashion by truncating $L_{{\rm synch},j+1}$
whenever it exceeds the local blackbody luminosity.  In the
Rayleigh-Jeans limit, the maximum luminosity that can exit shell $j+1$
at frequency $\nu$ is
$$
L_{{\rm max},j+1}(\nu)d\nu=2kT_{e,j+1}{\nu^2\over c^2}A_{j+1}^{\rm d} d\nu,
\qquad A_{j+1}^{\rm a}=2\pi l_{j+1}^2(2\cos\theta_{H,j+1}+
\sin^2\theta_{H,j+1}), \eqno (2.3.3) 
$$
where $A_{j+1}^{\rm a}$ is the surface area of shell $j+1$ (including the
polar caps).

Thus, we have the following recursion relation for calculating the
synchrotron luminosity exiting shell $j+1$:
$$
L_{{\rm synch},j+1}={\rm Min}\left[(1-P_{j,j+1}^{\rm aa})L_{{\rm synch},j}+j_{{\rm synch},j+1}
V_{j+1},L_{{\rm max},j+1}\right]\equiv L_{{\rm synch},j}+L_{s,j+1}.\eqno (2.3.4)
$$
Starting with $L_{{\rm synch},1}=0$, this recursion allows us to estimate
the synchrotron luminosity $L_{{\rm synch},j}$ exiting each succeeding shell
$j$ outwards from the center.  

The quantity $L_{s,j+1}$ on the right hand side of equation (2.3.4)
represents the net luminosity of shell $j+1$ alone (as distinct from
$L_{{\rm synch},j+1}$ which is the luminosity of all shells from 2 to
$j+1$).  The integral of $L_{s,j+1}d\nu$ gives the cooling of shell
$j+1$ as a result of synchrotron radiation.

\subsubsection{Compton Scattering}

Consider a photon of frequecy $\nu'$ that is Compton-scattered once by
thermal electrons at a temperature $T_e$.  Define
$C(\nu',\nu;T_e)d\nu$ to be the isotropically averaged probability
distribution of the frequency $\nu$ of the scattered photon.  This
distribution has been calculated by Jones (1964), with corrections
published by Coppi \& Blandford (1990).  The formulae given in these
papers are exact and include relativistic corrections corresponding to
the Klein-Nishina regime.

In each shell $j$, we define $L_{{\rm C,in},j}$ to be the net radiative
energy which is Compton-scattered per second, and $L_{{\rm C,out},j}$ to be
the net output luminosity in the scattered photons.  By the definition
of $C(\nu',\nu;T_e)$, we have
$$
L_{{\rm C,out},j}d\nu=\left[\int L_{{\rm C,in},j}(\nu')C(\nu',\nu;T_{e,j})d\nu'
\right]d\nu.
\eqno (2.3.5)
$$
This relation allows us to calculate the contribution of each shell to
Comptonization, provided we know $L_{{\rm C,in},j}$.  The total outgoing
luminosity from shell $j$, including the contributions from
synchrotron and bremsstrahlung emission, is thus
$$
L_{{\rm out},j}=L_{{\rm brem},j}+L_{s,j}+L_{{\rm C,out},j}.\eqno (2.3.6)
$$

To calculate $L_{{\rm C,in},j}$, we note that any radiation emitted in any
shell $k$ of the ADAF or ring $l$ of the thin disk has a certain
probability of being scattered in shell $j$, the probability being
given by the various matrices discussed in \S2.3.1.  Therefore,
$$
L_{{\rm C,in},j}=\Sigma_kP_{kj}^{\rm aa}L_{{\rm out},k}
          +\Sigma_lP_{lj}^{\rm da}L_{{\rm disk},l}, \eqno (2.3.7)
$$
where $L_{{\rm disk},l}$ is the spectral luminosity of ring $l$ of the thin
disk.  Equation (2.3.7) and eqs. (2.3.8) and (2.3.10) below close the
loop for calculating the effect of Comptonization.  The equations are,
however, strongly coupled among different shells and rings, and
require an iterative technique of solution.

\subsubsection{Outer Thin Accretion Disk}

The outer disk has two sources of energy: (i) its own internal viscous
dissipation, and (ii) irradiation from the ADAF and corona.  By the
standard equations of thin accretion disk theory (cf. Frank et
al. 1992), we have the following expression for the effective
temperature of ring $j$ of the disk:
$$
\sigma T_{{\rm eff},j}^4A_j^{\rm d} = {3GM\dot M\over2R_S}\left\{
{1\over l_{j-1}}\left[1-{2\over3}\left({r_{\rm tr}\over l_{j-1}}\right)^{1/2}\right]
-{1\over l_{j}}\left[1-{2\over3}\left({r_{\rm tr}\over l_{j}}\right)^{1/2}\right]
\right\}
                        +\Sigma_k P_{kj}^{\rm ad}L_{{\rm out},k},\eqno (2.3.8)
$$
where
$$
A_j^{\rm d}=4\pi(l_j^2-l_{j-1}^2) \eqno (2.3.9)
$$
is the surface area of the ring, counting both the top and bottom
surfaces.  The second term on the right of equation (2.3.8) is the net
luminosity absorbed by the ring via irradiation from the ADAF and
corona.

We assume that the disk emits as a blackbody.  Thus, the outgoing
luminosity from ring $j$ is written as
$$
L_{{\rm disk},j}d\nu={A_j^{\rm d}(2h\nu^3/c^2)d\nu
\over \exp(h\nu/kT_{{\rm eff},j})-1}.\eqno (2.3.10)
$$

Note that in our model we do not need to assume a specific value of
the viscosity parameter $\alpha_d$ in the outer disk.  All we require
is that the emission from the disk has a locally blackbody form, and
this is a valid approximation so long as $\alpha_d$ in the disk is
small enough to make the gas optically thick in the vertical
direction.

In our calculations, we assume that all the irradiating flux is
absorbed by the disk.  Actually, a fraction of the flux is reflected
and only the remainder is absorbed.  The error we make by ignoring
reflection is negligible since the contribution of the disk to the
calculated spectra is very small (\S3.2).

\subsubsection{Iterative Scattering}

The sychrotron and bremsstrahlung luminosities, $L_{s,j}$ and $L_{{\rm
brem},j}$, can be directly calculated once the electron temperatures
$T_{e,j}$ are given.  However, the Comptonized radiation is less
straightforward since the output from each shell acts as an input to
other shells.  A convenient technique to handle this situation is the
iterative scattering method or $\Lambda$ iteration method (Poutanen \&
Svensson 1996, Mihalas 1978), where the radiative transfer equation is
solved for each scattering order.  (The technique is very efficient at
low optical depths, as in the present work, but is less useful when
the optical depth is greater than a few.)  Operationally, for our
problem, the iterative scattering method consists of iterating on
equation (2.3.5) until convergence.  At the first iteration, we set
$L_{{\rm C,out},j}=0$ and calculate the various $L_{{\rm C,in},j}$ via
equations (2.3.6)--(2.3.10).  Using these $L_{{\rm C,in},j}$, we
calculate new estimates of $L_{{\rm C,out},j}$.  Then, we use these to
obtain new estimates of $L_{{\rm C,in},j}$, calculate new $L_{{\rm
C,out},j}$, etc. until convergence.

The net radiative cooling in each shell is given by
$$
Q_j^-=\int\left(L_{{\rm out},j} -L_{{\rm C,in},j}\right)d\nu. \eqno (2.3.11)
$$
This quantity is used in equation (2.2.6) to solve for the ion and
electron temperatures in the various shells.

Once the temperatures and spectra have converged, we obtain the spectral flux
$F(\nu)d\nu$ seen by an observer at distance $D$ as follows,
$$
4\pi D^2F(\nu)=\Sigma_jE_j^{\rm a}L_{{\rm out},j}
              +\Sigma_kE_k^{\rm d}L_{{\rm disk},k}.\eqno (2.3.12)
$$
The escape probabilities $E_j^{\rm a}$ and $E_k^{\rm d}$ correspond to the
particular inclination $i$ of the system, as described in \S2.3.1.

%####################### V404 Cyg in Quiescence #####################
\section{V404 Cyg in Quiescence}

We have applied the above model to the SXT source V404 Cyg in
quiescence.  \S 3.1 describes our analysis of ASCA observations of the
source in the X-ray band and \S3.2 describes observations in other
wavelength bands.  \S3.3 compares the observations with theoretical
model spectra and shows that the ADAF model provides an excellent fit
to the data.  Finally, \S3.4 describes a re-analysis of the source
A0620--00.

\subsection{Analysis of archival X-ray data}

V404 Cyg was observed by ASCA on 9-10 May 1994 for a total exposure
time of about 40 ksec with the GIS detectors and 35.5 ksec with the
SIS detectors. The results of these observations are not currently
available in the literature.  We therefore retrieved the data from the
HEASARC archives and performed a spectral analysis in order to test
the predictions of the ADAF model. The good data have been selected
following the procedure described in Day et al. (1995)

The spectral analysis was somewhat difficult because the source flux
is comparable to the background flux. Therefore, special care was
taken with the background subtraction. For the two GIS detectors, we
determined the background in two ways: using observations of blank
fields at high latitudes (Method I), and using data extracted from an
annular region around the source (Method II).  Given the low Galactic
latitude of the source (b = $-2.2^{\circ}$), unresolved emission from
the Galactic plane may contribute to the source flux; therefore we
prefer Method II for the GIS observations. (The results of our
analysis, however, show that Method I gives results that are fully
consistent with Method II).  For the GIS detectors we did the full
analysis using different extraction radii for the source region (4, 5
and 6 arcmin).  We found that the optimum radius (the one which
maximizes the signal to noise ratio) was 5 arcmin.

For the two SIS detectors, we determined the background using data
from high-latitude blank fields.  We used the November 1994 release of
the background files provided by the ASCA Guest Observer facility at
HEASARC. For these detectors we fixed the extraction radius for the
source region at 3.5 arcmin.  V404 Cyg is clearly detected in the SIS
detectors from about 0.7 to 8.5 keV with an average count rate of
$0.0102\pm0.0008$ \ctss.  The average count rate in the GIS detectors
is very similar: $0.0107\pm0.0015$ {\ctss} over the same energy band.
In order to use the $\chi^2$ statistic in the model fitting, we
grouped the raw SIS spectra so that each spectral bin contained at
least 20 photons (40 photons for the GIS spectra).  We found that the
spectra determined using the SIS0, SIS1, GIS2 and GIS3 detectors were
individually consistent with each other within the errors. Thus, we
combined the four individual spectra to obtain the final spectrum.

We fitted the data to a simple power law (PL) model with two free
parameters: a single photon index, $\alpha$, and a column density
N$_{\rm H}$.  This model with $\alpha$ = 2.1 and N$_{\rm H}$ = $1.1
\times 10^{22}$ cm$^{-2}$ provides an excellent fit to the ASCA data
($\chi^2_\nu = 1.0$ for 105 d.o.f; Table 1).  A comparably good fit is
provided by a thermal bremsstrahlung (TB) model with a temperature of
4.7 keV because the shapes of the TB and PL models are similar over
the energy range 0.7 - 8.5 keV.  A blackbody (BB) model provides a
poorer fit than either the PL or the TB model; nevertheless, a BB
model cannot be ruled out because the difference in $\chi^2_\nu$ is
not statistically significant.  (The probability that $\chi^2_\nu$ is
larger than the value derived from the PL or TB fit is 49\%, whereas
it is $\sim 15$ \% in the case of the BB fit.)  

An inspection of the residuals for the BB fit indicates, however, that
above 5 keV the data points lie systematically above the model;
moreover, interstellar absorption is clearly a dominant factor below
about 2 keV, which is near the peak of the BB model.  Consequently, we
re-fitted the data above 2 keV excluding the parameter N$_{\rm H}$ and
found that the BB model gives a poor description of the data:
$\chi^2_\nu$ = 1.4 (62 d.o.f.)  with a corresponding probability of
only 2\%.  On the other hand, the PL fit is still acceptable
($\chi^2_\nu =$ 1.1), and gives a photon index of
$1.68^{+0.27}_{-0.29}$, consistent with the value derived from the fit
between 0.7 and 8.5 keV.  Therefore, although the BB model cannot be
completely ruled out by the current data, these results suggest that
it is not the correct description of the source spectrum.

The best fit results for the three models are listed in Table 1, and
the combined GIS+SIS unfolded spectrum for the PL model is shown in
Figure 1.  In Figure 2a we show the allowed grid of variations of column
density N$_{\rm H}$ and photon index $\alpha$.  Assuming the PL model,
the 1-10 keV unabsorbed X-ray flux is $8.2 \times 10^{-13}$ \ergscm,
corresponding to a luminosity of $1.2 \times 10^{33}$ \ergs.

The error box shown in Figures 3--5 is given at the 2$\sigma$ level
and was derived as follows. The upper/lower curves of the box are
defined as the maximum/minimum fluxes allowed by any combinations of
$\alpha$ and its corresponding normalization on the 2$\sigma$ contour
in Figure 2b.  The box extends from 0.5 to 10 keV.  The ``throat'' of
the box represents the most accurately measured region of the
spectrum.  The center of the throat corresponds to: $E=3.5$ keV
(i.e. $\log\nu=17.93$), and $\nu F_\nu=3.56\times10^{-13} ~{\rm
ergs\,s^{-1}\,cm^{-2}}$ (i.e. $\log(\nu F_\nu)=-12.45$).  The models
described in \S3.2 were fitted to this flux measurement.

V404 Cyg was observed with the ROSAT satellite and the PSPC detector
by Wagner et al. (1994).  They fitted the data with a power-law model
and found a photon index of $\alpha\sim7$ (no uncertainty given).  NMY
re-analyzed these ROSAT data and found the following value for the
power-law index, $\alpha=4.0_{-1.5}^{+1.9}$, which is consistent with
the ASCA value given in Table 1.  The ASCA results reported here are
superior to the ROSAT results and supersede them primarily because
ASCA has a much greater useful bandwidth for observing V404 Cyg:
0.7--8.5 keV {\it vs.} 0.7--2.4 keV.

\subsection{Observations of V404 Cyg in Other Wavelength Bands}

The dereddened B, V and R magnitudes of V404 Cyg, and the fraction of
the total flux in these bands contributed by the accretion disk are
given by Casares et al. (1993).  From these data NMY extracted the
following flux estimates for the accretion disk in the B, V and R
bands: $\log(\nu F_{\nu}) = -11.256$, $-11.293$, $-11.368$,
respectively at $\log(\nu) = 14.834, ~14.737, ~14.632$ ($\nu$ in Hz,
$F_{\nu}$ in ${\rm ergs\,s^{-1}\,cm^{-2}\,Hz^{-1}}$).

V404 Cyg is significantly reddened ($A_V\approx4.0$ mag).  Therefore,
it is not easy to measure the UV flux.  However, following the
approach used by Marsh, Robinson \& Wood (1994) for A0620--00, we can
obtain a limit on the EUV flux.  The HeII 4686 emission line is absent
in the sensitive blue spectrum of V404 Cyg shown in Figure 1 (top) of
Casares et al. (1993).  A reddened version of this spectrum in digital
form was kindly provided to us by J. Casares.  We dereddened the
spectrum using $A_{\rm V}$ = 4.0 mag (Cardelli, Clayton \& Mathis
1989), which admirably flattened a 300 \AA\ interval of the continuum
centered on HeII 4686.  We assumed that the full width at zero
intensity (FWZI) of the $\lambda 4686$ line is 40 \AA\ (which is
approximately the FWZI of the H$_{\beta}$ emission line), and measured
the total intensities in three adjacent 40 \AA\ bands with the central
band centered on $\lambda 4686$. These three intensities differ by
somewhat less than the uncertainties in the individual bands, which we
estimated from the rms fluctuations in the (1.4~\AA-wide) pixels to be
1.2\%.  Using this value as an estimate of the standard deviation in
the intensity of a 40 \AA\ band, we obtain the following upper limit
on the equivalent width of the line: EW($\lambda 4686$) $<$ 1.4 \AA\
(3$\sigma$).

We use the absence of HeII 4686 to constrain the EUV flux from either an 
inner accretion disk or from an ADAF region.  (For details see the derivation 
for A0620--00 by Marsh et al. 1994). We assume that every photon in the
energy range 55 $<$ h$\nu$ $<$ 280 eV photoionizes HeI just once, and that a
fraction $\epsilon$ = 0.2 of the subsequent recombinations produce a
HeII~4686 photon.  The fraction $\alpha$ of the EUV photons that photoinize
HeI is the same as the fraction of the sky (as viewed from the compact object) 
that is covered by the accretion disk and by the secondary star, which we 
estimate to be 0.008 (Shahbaz et al. 1994) and $\approx$ 0.02 (Frank, King, 
\& Raine 1992), respectively.  Thus, we take $\alpha \approx$ 0.03.  Here 
we have made the reasonable assumption that both the annular disk in
our ADAF model and the complete thin disk stop about the same fraction
of the EUV photons.  The dereddened continuum flux at B is $f_{\nu}
\approx 3.4 \times 10^{-26}$ ergs s$^{-1}$ cm$^{-2}$ Hz$^{-1}$
(Casares et al. 1993).  Assuming an average EUV photon energy of
$\overline{\rm E}$ = 100 eV and D = 3.5 kpc, we can now translate the
limit on the HeII 4686 photon flux into a 3$\sigma$ limit on the EUV
luminosity:
$$ 
L_{\rm EUV} < 2 \overline{\rm E} \frac{\rm EW}{\lambda} \frac{f_{\nu}}{h}
\frac{4\pi D^{2}}{\alpha\epsilon} = 1.2 \times 10^{35} {\rm ergs}~{\rm s}^{-1}.
\eqno (3.1)
$$
The corresponding limit on the average flux density in the 55-280 eV
band is $\log(\nu F_{\nu})<-10.44$ (3 $\sigma$).

\subsection{Models of V404 Cyg in Quiescence}

V404 Cyg is a well-studied source and many of the system parameters
are reasonably well constrained.  We choose a black hole mass
$M=12M_\odot$ and inclination $i=56^o$ (Shahbaz et al. 1994).  Given
the size of the Roche lobe around the black hole, we estimate the
outer radius of the accretion flow to be $r_{\rm out}=10^5$.  The
H$\alpha$ line of V404 Cyg shows a maximum velocity of $1140 ~{\rm
km\,s^{-1}}$ (NMY; Casares et al. 1993), which corresponds to a
transition radius $\log(r_{\rm tr})=4.4$ (cf. eq. 2.1.3).  The mass
accretion rate $\dot M$ is not known.  In each of the models described
below, we have adjusted $\dot M$ so as to fit the X-ray flux in the
``throat'' of the ASCA error box, as described in \S3.1.  The fluxes
corresponding to the model spectra are calculated assuming a distance
$D=3.5$ kpc in equation (2.3.12) (Wagner et al. 1992, Shahbaz et
al. 1994).

This still leaves $\alpha$, $\beta$ and $\delta$ undetermined.  In
much of our previous work we have chosen values of $\alpha$ in the
range 0.1 to 1.  There are two reasons for this.  First, theoretical
models of the dwarf nova instability strongly suggest that accretion
disks in cataclysmic variables have low values of $\alpha\sim0.01$
when they are in a cool state and larger values $\sim0.1$ when they
switch to a hot state (Smak 1993, Cannizzo 1993, Mineshige \& Kusunose
1993).  Since our ADAFs are significantly hotter than the hottest
state of cataclysmic variables it seems reasonable to choose a value
of $\alpha$ which is correspondingly larger.  Secondly, ADAF models
provide a natural explanation (Narayan 1996b) for some spectral states
of black hole X-ray binaries, especially the ``low state,'' but the
models work only if $\alpha$ has a fairly large value.  For the
self-similar models considered by Narayan (1996b), the models required
$\alpha\rightarrow1$.  With the more realistic global solutions that
are now being developed (Narayan et al. 1997, Chen et al. 1997,
Nakamura et al. 1996, Abramowicz et al. 1996) and that are the basis
of the present paper, the requirement is a little less stringent:
$\alpha\sim 0.3$ perhaps.  We therefore select $\alpha=0.3$ for our
baseline model.

The parameters $\beta$ and $\delta$ are much more open, since there
are very few observational constraints on their values.  Fortunately,
the value of $\beta$ seems to have relatively little effect on the
results.  Here we try two values: $\beta=0.5$, $0.95$.  In the case of
$\delta$, our standard choice is $\delta=10^{-3}$ ($\sim m_e/m_p$).

Figure 3a shows two models of V404 Cyg corresponding to $\beta=0.5$
and $0.95$, and standard values for the other parameters.
Superimposed on the model spectra are the error box from the ASCA
observations (described in \S3.1) and three dots corresponding to the
optical data and an arrow corresponding to the EUV upper limit (see
\S3.2).  Table 2 lists the parameters we have chosen for these two
models (as well as other models described below), and Table 3 gives a
few key results of the models.  The spectra shown in Figure 3a
correspond to Model 1 and Model 2 in the Tables.  Model 1 is our
``standard model.''

Interestingly, the value of $\dot M$ which we derive by fitting the
X-ray flux is $\sim {\rm few}\times10^{-10}-10^{-9} M_\odot\,{\rm
yr^{-1}}$ for the various models presented in this paper.  This is
slightly smaller than, but comparable to the estimate of $\dot M$
obtained by King (1993) from evolutionary considerations.  Thus, in
our model, a reasonable fraction of the mass that is transfered by the
secondary star is immediately accreted via the inner ADAF.  However,
some fraction of the mass does appear to be stored in the outer disk
and it is this storage which presumably causes V404 Cyg to go
occasionally into outburst (Mineshige \& Wheeler 1989).

Both models in Figure 3a have hard spectra in the ASCA band with
photon indices $\alpha_N\sim2-2.2$.  These spectra are in excellent
agreement with the observed spectral constraints, thus impressively
confirming a key prediction of the NMY model.  Note that the
calculated spectra shown here are slightly softer than those obtained
by NMY.  The difference is a result of the more careful modeling done
here, especially the use of a global flow instead of the self-similar
solution employed earlier, and a better treatment of Comptonization.

Figure 3b shows two thin disk models (with no ADAF or corona) which
have been adjusted to fit the optical flux of V404 Cyg.  The solid
line corresponds to a model in which the accretion rate in the disk
$\dot M_d$ is assumed to be constant as a function of radius.  The fit
to the data gives $\dot M_d= 8\times10^{-10} ~M_\odot{\rm yr^{-1}}$.
Note the huge flux that this model predicts in EUV and soft X-rays and
the lack of emission in harder X-rays.  Both features are inconsistent
with the observations.  The dashed line in Figure 3b corresponds to
another model in which the mass accretion rate is assumed to vary as
$\dot M_d=\dot M_{\rm out}(r/r_{\rm tr})^3$.  This model has a nearly
constant effective temperature as a function of radius and is more in
line with what is expected for a quiescent thin disk which is on the
cool branch of the standard S-curve (cf. Mineshige \& Wheeler 1989).
In this case, a fit to the optical flux is obtained with $\dot M_{\rm
out}=5\times10^{-8} ~M_\odot {\rm yr^{-1}}$.  This model does not
produce any radiation in bands harder than the optical and is clearly
inconsistent with the ASCA observations.  Thus we conclude that, just
as in A0620--00 (NMY), the data on V404 Cyg cannot be fit with any
model that is composed purely of a thin accretion disk.

In the models shown in Figure 3a, nearly all the flux is from the ADAF
(see Table 2).  The peak in the optical/UV region of the spectrum is
due to synchrotron emission, the bump in the spectrum at EUV
wavelengths is due primarily to singly Compton-scattered photons, and
the emission in the ASCA band is mostly doubly Compton-scattered
radiation, along with some bremsstrahlung emission.  (The bulk of the
bremsstrahlung is at yet higher photon energies $\sim100$ keV.)  The
outer thin disk produces very little radiation in these models.  This
is because we have used a very large value for the transition radius
$r_{\rm tr}$ (selected on the basis of the H$\alpha$ line width).
Consequently, even though the thin disk is a much more efficient
radiator than the ADAF, the fact that its energy budget is so much
smaller (by a factor $\sim r_{\rm tr}^{-1}=10^{-4.4}$) means that its
emission is quite negligible.  A small amount ($\sim10\%$) of the flux
at infrared wavelengths is in fact from the disk, but it is not
visible as a separate peak.  The models presented here differ in this
respect from those discussed by NMY where $r_{\rm tr}$ was smaller
($\sim10^3$) and therefore the outer disk was a more important
contributor to the optical flux.

A noteworthy feature of the models shown here is that by taking
standard parameter values and adjusting just one parameter, viz. the
value of $\dot M$, we are able to fit both the X-ray and optical
fluxes, as well as the spectral shapes in these two bands.  It appears
that the $\beta=0.5$ model fits the optical data better than does the
$\beta=0.95$ model, but we do not believe that the difference is
significant at this point.  As we mentioned in \S2, the models still
involve approximations, some of which may modify the results somewhat;
the use of non-relativistic instead of relativistic physics is a case
in point.  In view of this, we consider both models in Figure 1a to be
satisfactory.

How sensitive are the results to the choice of parameter values?  The
answer is: extraordinarily little.

Figure 4a shows three models where we have kept all parameters the
same as for the $\beta=0.5$ model in Figure 1a, except the black hole
mass, which we varied: $M=8M_\odot, ~12M_\odot, ~16M_\odot$.  (These
correspond to Models 3, 1, and 4, respectively, in Tables 2 and 3.)
All three models fit the data very well.  The synchrotron peak moves
to slightly higher energies with decreasing mass.  This is understood
by noting that the electron temperatures in the three models are
essentially the same (Table 3), and so the synchrotron peak is
determined primarily by the magnetic field strength $B$.  As shown in
NMY, the field strength at a given radius $r$ scales with mass as
$B\propto M^{-1/2}$, which explains the trend seen in Figure 4a.

Figure 4b shows four models (Models 5, 1, 6, 7 in Tables 2 and 3)
where the transition radius is varied: $\log r_{\rm tr}=4.9, ~4.4$,
$3.9, ~3.4$.  The radiation from the ADAF is virtually unaffected by
the change in $r_{\rm tr}$.  However, as $r_{\rm tr}$ decreases, the blackbody
emission from the outer disk becomes more important and this leads to
enhanced flux in the optical band.  We consider all the models in
Figure 4b to be acceptable fits to the data.

We have also tried varying the inclination angle $i$ in the standard
model, but the differences in the model spectra are so small that the
spectra lie practically on top of each other.  The reason for this is
two-fold.  First, the ADAF is nearly spherically symmetric and
optically very thin.  Its emission is therefore almost precisely
isotropic.  The flux from the outer disk does depend strongly on the
inclination, varying as $\cos i$.  However, the total flux from the
disk is so small (see Table 3) that even the strong $\cos i$ variation
leads to almost no detectable effect.

For the reasons discussed above, the exact manner in which we model
the outer disk is not important for the spectral fits presented here.
For instance, we have modeled the run of $\dot M$ in the disk and the
corona somewhat arbitrarily according to equation (2.1.2).  However,
we could have used quite different models and the results would hardly
have changed.  In fact, even the choice of a constant total $\dot M$
between the disk and the corona is unimportant.  We could have made
$\dot M$ increase outward, as expected for a quiescent disk (Mineshige
\& Wheeler 1989), without affecting the spectral fit.

Figure 5a shows five models in which $\alpha$ is varied: $\alpha=0.1$,
0.2, 0.3, 0.4, 0.5.  Once again we see that the changes are minimal,
and all five models are consistent with the data.

Finally, Figure 5b shows the effect of varying the parameter $\delta$.
Four models are shown, corresponding to $\delta=0.001, ~0.00316,
~0.01, ~0.0316$.  Here we see that the model spectra do show modest
variations.  Increasing $\delta$ corresponds to increasing the direct
heating of the electrons, which leads to a higher equilibrium electron
temperature (Table 3).  This pushes all peaks to higher energies.  The
effect is sufficiently strong that for $\delta=0.0316$, the first
Compton-scattering peak moves into the ASCA band and is seen as a
steep component below about 1 keV.  We consider all four models to be
consistent with the present data; however, using better data it may be
possible to constrain the value of $\delta$.  We must caution the
reader once again that we need to develop a relativistic version of
the code before making such detailed comparisons.

\subsection{Re-Analysis of A0620--00 in Quiescence}

Figure 6 shows the various observational constraints on A0620--00, as
given by NMY.  Relative to V404 Cyg, we see that there is
significantly more information in the optical and UV bands; in fact,
the data show clear evidence for a downturn of the flux in the UV,
which is an important constraint on the model.  The EUV flux
constraint is also tighter compared to V404 Cyg.  The X-ray data are,
however, of poorer quality since the entire signal consists of only
$39\pm8$ net photons from a $3\times10^4$ s observation with ROSAT
(McClintock et al. 1995).  We rebinned the 34 PHA channels into 6
channels.  Ignoring the lowest channel, we performed a $\chi^{2}$
analysis similar to the one described in Sect. 3.1 with one important
difference: here we fixed the interstellar column density at N$_{\rm
H}$=$1.6 \times 10^{21}$ whereas in the case of V404 Cyg we determined
N$_{\rm H}$ using the ASCA data.  For $D = 1.0$ kpc and an energy
range of $0.4-2.5$ keV, we find $\alpha_{N} = 3.5 \pm 0.7$ (
$1\sigma$) and $L_{\rm X} \approx 5\times10^{30}$ ergs~s$^{-1}$, which
is in close agreement with our earlier results obtained using
maximum-likelihood fitting (NMY).  Using the coarsely binned data, we
computed the $2\sigma$ X-ray error box shown in Figures 6-7.  Note
that even though the column density has been fixed, the error box is
much wider than it is for V404 Cyg.

The throat of the X-ray error box corresponds to: $\log(\nu)=17.383$,
$\log(\nu F_\nu)=-13.789$.  As before, we have adjusted $\dot m$ in
each model so as to fit this flux.  The various other parameters of
A0620--00 are reasonably well-known.  Two system inclinations are
discussed in the literature, $i=70^o, ~40^o$ (see NMY for references),
with corresponding black hole masses of $M=4.4M_\odot, ~12M_\odot$.
Barret, McClintock \& Grindlay (1996) suggest an intermediate value:
$i=55^o$, $M=6.1M_\odot$, which we adopt for our standard model.  The
maximum velocity from the $H\alpha$ line is $2100~{\rm km\,s^{-1}}$,
which gives $\log(r_{\rm tr})=4.0, ~3.6, ~3.8$ for the three
inclinations mentioned above.  We take the distance to the source to
be 1.0 kpc (NMY).

Figures 6, 7 and Tables 4, 5 give results corresponding to a series of
models of A0620--00.  In general we see that the models agree fairly
well with the data.  The spectral slope in the X-ray band is
compatible with the $2\sigma$ range allowed by the observations,
especially when one considers that we fixed N$_{\rm H}$ in the
analysis of the data and thereby underestimated the uncertainties.
The predicted flux in the optical band is higher than the observed
flux by a factor of about 2 or 3.  As already discussed in \S3.3, the
model has several residual uncertainties and so a discrepancy of this
magnitude is not unreasonable.  Interestingly, all the model spectra
have downturns in the UV band, in good agreement with the
observations.  This is quite impressive when one recalls that in each
model we adjust only one parameter, viz. $\dot m$, and we do not use
the optical/UV data at all for determining this parameter.

We view the good agreement in the optical/UV region of the spectrum as
a confirmation of the model.  The optical/UV flux in the present
models is almost entirely contributed by synchrotron emission from the
ADAF.  This is in contrast to the models presented in NMY, where the
flux was primarily from the outer thin disk.  Indeed, in those models,
the optical region of the spectrum was fitted by adjusting the
transition radius $r_{\rm tr}$.  This additional degree of freedom has
been eliminated in the present models.

Figure 6a shows the effect of varying $\beta$.  For $\beta=0.5$ (our
standard model), the predicted optical flux is a little high but the
shape of the spectrum is in good agreement with the data; the
predicted colors agree reasonably well with the observations.  The
$\beta=0.95$ model, on the other hand, has the correct flux but is too
red.  Figures 6b, 7a, 7b show the effect of varying the black hole
mass, the transition radius, and the electron heating parameter
$\delta$.  Larger values of $\delta$ lead to better agreement with the
X-ray data, but since the X-ray error box is fairly uncertain we feel
that it would be premature to draw any conclusion from this.

%######################### Summary and Discussion ##################
\section{Summary and Discussion}

The principal result of this paper is that model spectra of the SXT
source V404 Cyg in quiescence are in very good agreement with
observations.  The model we consider is similar to that proposed by
NMY and consists of an ADAF over a wide range of radius from the black
hole horizon out to a transition radius $r_{\rm tr}>10^4$
Schwarzschild radii, and a thin accretion disk beyond the transition
radius.  Nearly all the observed radiation comes from the ADAF, the
flux from the outer thin disk being quite negligible.  By taking
standard values for various parameters (\S3.2, Tables 2, 3) and
adjusting only the mass accretion rate $\dot M$, we are able to
explain all the available data, namely the optical and X-ray fluxes,
the spectral shapes in these two bands, and an upper limit in EUV.  In
particular, we confirm the prediction of NMY that the X-ray spectrum
of V404 Cyg should be hard, with a photon index of order 2.  Our
analysis of archival ASCA data (\S3.1, Figure 1, Table 1) shows that
V404 Cyg has a photon index in the range 1.8 to 2.6 (90\% confidence
limits), which is perfectly consistent with the model prediction.
Moreover, the agreement between the observations and the model is
extremely robust in the sense that we have tried changing all the
parameters by fairly large amounts and the fit remains good (Figures
3--5).  In contrast, we find that it is impossible to fit the
observations with any model that is based only on a standard thin
accretion disk (see Figure 3b).  These results provide a strong
argument in favor of the advection-dominated accretion paradigm for
quiescent black hole SXTs.

We must emphasize that the quality of the X-ray data that the present
models of V404 Cyg are being tested against is much superior to the
X-ray data which NMY had to work with in the case of A0620--00.  For
A0620--00, McClintock et al. (1995) detected a total of only $39\pm8$
photons and there is a question even if all the photons came from the
accretion flow (some could have come from the secondary).  Even
assuming that the signal is entirely from accretion, one can obtain an
estimate of the photon index only by assuming a value for the hydrogen
column (based on radio data), and the photon index is only moderately
constrained (see the error box in Figure 6).  In contrast, the ASCA data
we have analyzed here correspond to $\sim1500$ photons, and the error
boxes we show in Figures 3--5 represent full $2\sigma$ limits even
after fitting for N$_{\rm H}$ with the same data.  The contrast in the
quality of data is obvious, and the present test of the model on V404
Cyg is significantly more stringent than the one presented in NMY.

The calculation techniques employed in this paper are also an
improvement over those used by NMY.  We use a global solution for the
flow dynamics (taken from Narayan et al. 1997 and Chen et al.  1997),
instead of the simpler self-similar solution employed by NMY.  We
allow for non-sphericity of the ADAF (Appendix A).  We also allow for
full coupling among all regions of the optically thin flow in
computing the Comptonization (we use over 1000 rays emanating from
each radial shell in calculating the scattering probabilities, see
\S2.3.1), instead of the radial one-dimensional calculation used in
NMY.

In view of these improvements we believe that the model spectra
calculated in this paper are fairly realistic.  In particular, we
believe the predicted spectral slope in the X-ray band is robust.  As
the calculations reported by Narayan (1996b) show, ADAFs have a
characteristic evolution of spectral shape as a function of increasing
$\dot M$.  Given the luminosity in the X-ray band, expressed in
Eddington units, there is generally an unambiguous X-ray spectral
index associated with that luminosity (see also Mahadevan 1996).  In
the case of V404 Cyg, we know that the mass of the black hole lies in
the range $10M_\odot$ to $15M_\odot$ (Shahbaz et al. 1994), and from
this we find that the X-ray luminosity in quiescence is $\sim 10^{-6}$
times the Eddington luminosity.  This immediately constrains the slope
of the spectrum to be $\alpha\ \sgreat\ 2$.  It is because of the
robustness of this result that our model fit remains good despite
changing the various parameters of the model by large amounts (Figures
3--5).

Further improvements to the model are, of course, possible and are
clearly desirable in view of the success we have had.  Principal among
the improvements we seek is the inclusion of relativistic effects in
the flow dynamics and radiative transfer.  Abramowicz et al. (1996)
have already obtained consistent relativistic flow solutions in Kerr
geometry.  What is needed now is to couple those solutions with
radiative transfer calculations which include the effects of
gravitational redshift, Doppler boosts and ray deflections, effects
which have been neglected in the present code.  It would also be
preferable to avoid some of the minor approximations made in the
present treatment of Comptonization (see \S2.3).  Finally, one might
wish to go beyond the thermal models considered here and allow for
deviations of the electron distribution function from a pure
Maxwellian form.

Among all the parameters of our model, the only one which produces any
significant variation in the spectrum is $\delta$.  This parameter
specifies the fraction of the viscous heating which goes directly into
electrons.  The default value of $\delta$ is the ratio of electron to
proton mass $\sim10^{-3}$, but we have also tried larger values of up
to a few per cent (Figures 5b, 7b).  We find that as $\delta$
increases, the equilibrium electron temperature goes up and this
causes the peaks in the spectrum to move to higher energies.  While
the current data are not yet sensitive enough to distinguish among the
various models, better data may be able to constrain the value of
$\delta$.  This would be extremely useful since our understanding of
viscous heating in magnetized plasmas is quite primitive at this point
and it is hard to imagine estimating $\delta$ with any assurance from
first principles.  Figures 3a and 6a show that the spectrum also has a
modest dependence on the magnetic field parameter $\beta$.  In
principle, future observations may be able to measure $\beta$, which
is another parameter that is hard to calculate purely from theory.

The models presented in this paper differ in one important respect
from those discussed by NMY.  Here, the transition between the ADAF
and the outer disk occurs at a large radius, $r_{\rm tr}=10^{4.4}$ in
the case of V404 Cyg and $r_{\rm tr}=10^{3.8}$ in A0620--00, whereas
NMY had $r_{\rm tr}\sim10^{3}$.  Consequently, the spectra calculated
in this paper are dominated entirely by the ADAF; even in the
infrared, where the disk radiates most of its flux, the disk emission
is only on the order of 10\% of the total flux in V404 Cyg and 25\% in
A0620-00.  (The exceptions are Model 7 in Tables 2, 3 and Models 5 and
6 in Tables 4, 5, where the disk radiates a larger fraction of the
infrared and optical luminosity, but these models are ruled out on
other grounds as discussed below.)  The models of NMY, on the other
hand, were dominated by emission from the outer disk.  The fact that
the outer disk is completely unimportant to the spectral fits
presented here means that the exact details of our model of the disk
are not relevant.  We have for simplicity assumed that the total $\dot
M$ in the disk plus the corona is independent of $r$ and we have
adopted a specific form for the relative $\dot M$ in the two
components (eq. 2.1.2).  In fact, SXTs in quiescence are expected to
have $\dot M$ varying with $r$ over the thin disk (cf. Mineshige \&
Wheeler 1989), but the error we make by ignoring this effect is very
small.

In addition, the fact that $r_{\rm tr}$ is large means that the new
models are also immune to a stability problem which affected the older
models (Wheeler 1996).  The issue here is the stability of the outer
thin disk.  (Recall that the ADAF itself is stable, Abramowicz et
al. 1995, Narayan \& Yi 1995b, Kato et al. 1996.)  It is well-known
that thin disks with effective temperatures $T_{\rm eff}$ $\sles\
10^4$ K have a characteristic S-shaped dependence in the $T_{\rm
eff}$-$\Sigma$ plane, where $\Sigma$ is the surface density.  The
S-curve arises because of the effect of hydrogen ionization on the
opacity.  Disks are stable only if they are either on the top or
bottom segment of the S, but are unstable if they are on the middle
segment.  The problem with the NMY models was that they had effective
temperatures which spanned (as a function of radius) a range of values
from $T_{\rm eff}<10^3$ K on the outside to $T_{\rm eff}\sim10^4$ K at
the transition radius.  This means that some portion of the disk in
those models is on the unstable middle branch (Wheeler 1996).

Since the present models have a much larger $r_{\rm tr}$, they have a
lower maximum effective temperature, $\sim1500-2000$ K in the case of
V404 Cyg (see Table 3) and $\sim3000-4000$ K in A0620--00 (Table 5).
These models are therefore safely on the lower stable branch of the
S-curve at all radii, thus eliminating Wheeler's objection.  In fact,
the models also eliminate another problem highlighted by Wheeler.  A
stable thin accretion disk in its low state (lower branch of the S)
has its mass accretion rate decreasing rapidly with radius, roughly as
$r^3$ (cf. the discussion in \S3.3 in connection with Figure 3b).  In
the NMY models, the inner edge of the disk is at such a small radius
that $\dot M$ at the inner edge, assuming that the disk is on the
lower branch of the S-curve, is much smaller than the $\dot M$ needed
to supply the ADAF.  This is no longer an issue for the models
presented here.  At the large transition radii we are considering now,
the disk easily supplies the $\dot M$ of the ADAF.

In this connection we should mention that $r_{\rm tr}$ is not very
well constrained by the observations.  Recall that we have estimated
$r_{\rm tr}$ on the basis of the H$\alpha$ line width (\S3.2).  From
this measurement we obtained the maximum velocity $v_{\rm max}$ of the
disk material and we then employed equation (2.1.3) to calculate
$r_{\rm tr}$.  However, observationally it is quite difficult to
measure $v_{\rm max}$ and the estimate we have used should in reality
be considered only a lower bound.  This means that our estimate of
$r_{\rm tr}$ is actually an upper bound, as far as the H$\alpha$
observations are concerned.  In addition, we show in Figure 4b and
Figure 7a that we can vary $r_{\rm tr}$ by a modest amount without
affecting the spectral fit.  Thus, the spectrum does not constrain
$r_{\rm tr}$ either, and there is no useful lower limit on $r_{tr}$
from the observations.

We might be able to use theoretical arguments, however, to derive a
lower bound on $r_{\rm tr}$.  For instance, we could turn Wheeler's
argument around and insist that $r_{\rm tr}$ must be such that the
thin disk does not cross over from the stable lower branch of the
S-curve to the unstable middle branch.  Tables 3 and 5 give the
maximum effective temperatures of the outer disks in the various
models presented here.  Consider Models 6 and 7 of V404 Cyg in Table
3, which have smaller values of $r_{\rm tr}$ than our standard Model
1.  Model 6 has a maximum effective temperature of 4090 K, which
allows the disk to lie within the lower branch of the S-curve.  This
model is therefore consistent.  Model 7 is, however, much too hot and
is clearly ruled out since a substantial part of the disk will be in
the unstable branch.  Thus we can say that, within the context of the
parameterisation given in equation (2.1.2), $\log(r_{\rm tr})$ in V404
Cyg must lie in the range $\sim4-4.5$.  (Other parameterisations will
give different lower limits.)  In the case of A0620--00, Table 5 shows
that $\log(r_{\rm tr})=3.8$ is acceptable, but $\log(r_{\rm tr})=3.4$
and 3.0 have too high an effective temperature in the outer disk to be
stable.

Another theoretical constraint comes from the requirement that the
outer disk should be able to undergo a limit cycle instability, since
this is believed to be the mechanism that produces the observed
outbursts of SXTs (Mineshige \& Wheeler 1989).  If $r_{tr}$ is too
large, the outer disk may become permanently stable and there would be
no limit cycle.  To determine the exact limit on $r_{tr}$ from this
argument we require time-dependent models of the outer disk, coupled
with a detailed model of the ADAF interior.  Such models are yet to be
constructed.

The models presented here have a very unique geometry and make
specific predictions which could be tested with future observations.
Except for the H$\alpha$ line emission and a small fraction of the
infrared flux, virtually all the observed radiation comes from within
a few tens of Schwarzschild radii of the black hole, i.e $R\
\sles\ 10^8$ cm.  If an eclipsing SXT is discovered (which would appear
to be just a matter of time considering the rate at which new SXTs are
being found), this prediction can be easily tested.  Direct
confirmation that the optical and UV flux come from a compact volume
around the black hole would be strong verification of the model since
it is difficult to imagine any competing model having this feature.

In addition, since the optical emission is from an ultra-hot ADAF, the
optical flux could exhibit fairly interesting variability.  It is not
possible to predict the variability amplitude at this time, but the
time scale can be easily estimated.  For an ADAF with a large value of
$\alpha\sim1$ (as in our models), the dynamical, thermal and viscous
time scales are all nearly equal:
$t\sim2\pi(GM/R^3)^{-1/2}=0.03\;(M/10M_\odot)\;(r/10)^{3/2}$ s.  We
thus expect quite rapid variations from the hot gas at $r\ \sles\
100$.  Detection of this variability would be a strong confirmation of
the model since it is inconceivable that any thin disk model would
produce such rapid changes.  In addition, we may also expect some
slower variations associated with the transition radius, since this is
where the mass supply to the ADAF originates; the dynamical time scale
at $r_{\rm tr}$ is an hour in the case of V404 Cyg and a few minutes
for A0620--00.

Yet another possibility is that the optical synchrotron radiation may
be polarized.  The model assumes for simplicity that the magnetic
field in the ADAF zone is isotropically tangled, but if the field has
any residual anisotropy it might lead to polarized emission.  Note,
however, that the synchrotron emission is highly self-absorbed; it is
not clear how much polarization is expected under these circumstances.

Quiescent SXTs and other accreting black holes with anomalously low
luminosities may be among the best systems available for ``proving''
the existence of event horizons in black holes (NMY).  Recall that the
bulk of the energy in an ADAF is retained as thermal energy of the gas
and is advected into the central star.  This is especially obvious for
the models presented in this paper, where less than 0.1\% of the rest
mass energy of the accreting gas is radiated (see the last column of
Table 3).  If the central star is a black hole with a true horizon,
the advected energy disappears completely and does not contribute in
any way to the observed spectrum.  On the other hand, if the star is
not a black hole, but a normal star with a surface, then the accreted
thermal energy would ultimately be reradiated from the surface and
would in fact dominate the spectrum.  The successful application of
the ADAF model to A0620--00 was used by NMY as an argument in favor of
the black hole nature of that source.  The fact that a similar model
is now found to work even more impressively in V404 Cyg strengthens
the argument considerably.

What can be done to make this ``proof'' of black hole horizons more
compelling?  We feel that work needs to be done on several fronts:

\noindent
1. The key feature of both A0620--00 and V404 Cyg in quiescence is that
they are systems with low $\dot M$.  This is consistent with the
theoretical result that optically thin ADAFs occur only at low $\dot
M$ (Abramowicz et al. 1995, Narayan \& Yi 1995b).  Theory also shows
that ADAFs become progressively more advection-dominated with
decreasing $\dot M$ (see Figure 11 in Narayan \& Yi 1995b).  We thus
expect the most massive advection, and therefore the strongest
evidence for the disappearance of thermal energy, at the lowest mass
accretion rates.  Substantial progress has already been made to test
this prediction.  The ADAF model has been successfully applied to the
low-luminosity Galactic Center source Sgr A$^*$ (Narayan, Yi \&
Mahadevan 1995), to the low-luminosity nucleus of the liner galaxy NGC
4258 (Lasota et al. 1996a), and to a more general class of
low-luminosity galactic nuclei (Fabian \& Rees 1995).  Independent
estimates of $\dot M$ for several of these sources strongly suggest
that their luminosities are far below what is expected for accretion
via a standard thin disk, which converts the usual 10\% of the rest
mass of the accreting material into radiation.  In other words, there
is clear circumstantial evidence that all these systems advect large
amounts of energy into their central stars without any re-radiation.
In view of the success of the ADAF paradigm in low-luminosity black
holes, perhaps it is time to switch the argument around and ask: Is
there any counter-example to the ADAF paradigm among black hole
candidates which accrete at a low rate: $\dot M<10^{-3}-10^{-2}$ in
Eddington units?  In other words, is there any low $\dot M$ black hole
which can be shown {\it not} to have an ADAF?

\noindent
2. It is necessary to extend the models to higher mass accretion
rates.  What happens as $\dot M$ increases?  Up to what value of $\dot
M$ do ADAF solutions survive, and what do the brighter ADAF systems
look like?  How do accretion flows switch from an ADAF to a standard
thin accretion disk when $\dot M$ becomes too large?  Some preliminary
answers to these questions were presented by Narayan (1996b), who
explained the ``low state'' of black hole X-ray binaries in terms of
advection-dominated accretion; more detailed work on the outburst
of Nova Muscae 1991 is reported by Narayan \& McClintock (1997).
Higher $\dot M$ systems are much brighter and easier to observe, and
there is a large database on such systems, both among X-ray binaries
and active galactic nuclei.  There is thus enormous scope for testing
the ADAF paradigm in these systems.  In this connection, Yi (1996) has
applied ADAF models to try and explain quasar evolution.

\noindent
3. Thirdly, we feel that it is critical to demonstrate that there is a
clear difference between ADAF accretion on to a black hole and ADAF
accretion on to a normal star.  As already mentioned, a star with a
surface will re-radiate the accreted thermal energy and will therefore
be significantly more luminous than a similar black hole system.  The
spectra too will presumably differ.  There are several opportunities
for exploring these differences.  Many cataclysmic variables with low
$\dot M$ appear to resemble SXTs in the sense of having a truncated
outer thin disk and a central ADAF (e.g. Meyer \& Meyer-Hofmeister
1994).  The interaction between the ADAF and the central white dwarf
in these binaries is a topic which has hardly been explored.  An even
better opportunity is presented by SXTs such as Cen X-4 and Aql X-1
which are similar to black hole SXTs in many respects, but are known
to consist of accreting neutron stars by the fact that they have Type
1 X-ray bursts.  Indeed, as NMY pointed out, Cen X-4 in quiescence is
nearly a hundred times brighter than A0620--00 (a black hole system) in
quiescence; this is consistent with the ADAF paradigm.  A satisfactory
understanding of the luminosities and spectra of neutron star SXTs,
and how these differ from the corresponding characteristics of black
hole SXTs, would go a long way toward a definitive ``proof'' of black
hole horizons.  Tanaka \& Shibazaki (1996) emphasize some difficulties
in reconciling ADAF models with the observed spectra of neutron star
SXTs.

\noindent
4. Finally, it is possible that an accreting star could get rid of the
thermal energy in the accretion flow, not by swallowing it through an
event horizon, but by ejecting it via an outflow or jet.  To eliminate
this loophole it is necessary to show that black hole candidates with
ADAFs do not have sufficiently strong energy outflow via jets.  We
cannot at present judge how feasible this will be.

\bigskip\noindent
Acknowledgements: We thank G. Rybicki and J. Poutanen for helpful
discussions on Comptonization, K. Ebisawa, Y. Tanaka and K. Terada for
advice on the analysis of ASCA data, J. Casares for providing an
optical spectrum of V404 Cyg, M. Garcia and I. Yi for comments on the
manuscript, and an anonymous referee for useful suggestions.  This
research has made use of data obtained through the HEASARC Online
Service, provided by the NASA/Goddard Space Flight Center.  R.N. was
supported in part by NASA grant NAG 5-2837.  Partial support for
J.E.M. was provided by the Smithsonian Institution Scholarly Studies
Program.

\newpage
{
\footnotesize
\StartRef
\noindent {\large \bf References} \\
\Ref Abramowicz, M. A., Chen, X., Grantham, Lasota, J.-P. 1996, ApJ,
in press (astro-ph/9607021) \\
\Ref Abramowicz, M. A., Chen, X., Kato, S., Lasota, J.-P, \& Regev, O. 1995,
ApJ, 438, L37 \\
\Ref Abramowicz, M. A., Czerny, B., Lasota, J.-P, \& Suszkiewicz, E. 1988,
ApJ, 332, 646 \\
\Ref Abramowicz, M. A., Jaroszy\'nski, M., \& Sikora, M. 1978, A\&A, 
63, 221 \\
\Ref Barret, D., McClintock, J. E., Grindlay, J. E. 1996, ApJ, 473, 963 \\
\Ref Begelman, M. C. 1978, MNRAS, 184, 53 \\
\Ref Bjornsson, G., Abramowicz, M. A., Chen, X., \& Lasota, J.-P.
1996, ApJ, 467, 99 \\
\Ref Blandford, R. D., \& Znajek, R. 1977, MNRAS, 179, 433 \\
\Ref Cannizzo, J. K. 1993, in Accretion Disks in Compact Stellar Systems,
ed. J. C. Wheeler (Singapore: World Scientific), 6 \\
\Ref Cardelli, J. A., Clayton, C. C., \& Mathis, J. S. 1989, ApJ, 345, 245 \\
\Ref Casares, J., Charles, P. A., Naylor, T., \& Pavlenko, E. P. 1993,
MNRAS, 265, 834 \\
\Ref Chen, X. 1995, MNRAS, 275, 641 \\
\Ref Chen, X., Abramowicz, M. A., \& Lasota, J.-P. 1997, ApJ, in press 
(astro-ph/9607020) \\
\Ref Chen, X., Abramowicz, M. A., Lasota, J.-P., Narayan, R., \& Yi, I. 1995,
ApJ, 443, L61 \\
\Ref Coppi, P. S. \& Blandford, R. D. 1990, ApJ, 245, 453 \\
\Ref Day et al. 1995, ``The ABC Guide to ASCA Data Reduction'' \\
\Ref Esin, A. A. 1996, ApJ, submitted \\
\Ref Fabian, A. C. \& Rees, M. J. 1995, MNRAS, 277, L5 \\
\Ref Fishbone, L. G., \& Moncrief, V. 1976, ApJ, 207, 962 \\
\Ref Frank, J., King, A., \& Raine, D. 1992, Accretion Power in
Astrophysics (Cambridge: Cambridge Univ. Press) \\
\Ref Hawley, J. 1996, in Physics of Accretion Disks, eds. S. Kato, 
S. Inagaki, S. Mineshige, J. Fukue (Gordon and Breach) \\
\Ref Honma, F. 1996, PASJ, 48, 77 \\
\Ref Jones, F. C. 1968, ApJ, 167, 1159 \\
\Ref Kato, S., Abramowicz, M. A., \& Chen, X. 1996, PASJ, 48, 67 \\
\Ref Katz, J. 1977, ApJ, 215, 265 \\
\Ref King, A. R. 1993, MNRAS, 260, L5 \\
\Ref Kusunose, M., \& Mineshige, S. 1996, ApJ, 468, 330 \\
\Ref Lasota, J.-P., Abramowicz, M. A., Chen, X., Krolik, J., Narayan, R.,
\& Yi, I. 1996a, ApJ, 462, 142 \\
\Ref Lasota, J.-P., Narayan, R., \& Yi, I. 1996b, A\&A, in press 
(astro-ph/9605011) \\
\Ref Mahadevan, R. 1996, ApJ, in press (astro-ph/9609107) \\
\Ref Mahadevan, R., Narayan, R., \& Yi, I. 1996, ApJ, 465, 327 \\
\Ref Marsh, T. R., Robinson, E. L., \& Wood, J. H. 1994, MNRAS, 266, 137 \\
\Ref McClintock, J. E., Horne, K., \& Remillard, R. A. 1995, ApJ, 442, 358 \\
\Ref Meyer, F., \& Meyer-Hofmeister, E. 1994, A\&A, 288, 175 \\
\Ref Mihalas, 1978, Stellar Atmospheres (San Francisco: Freeman) \\
\Ref Mineshige, S. \& Kusunose, M. 1993, in Accretion Disks in Compact
Stellar Systems, ed. J. C. Wheeler (Singapore: World Scientific), 370 \\
\Ref Mineshige, S., \& Wheeler, J. C. 1989, ApJ, 343, 241 \\
\Ref Nakamura, K., Matsumoto, R., Kusunose, M., \& Kato, S. 1996,
PASJ, 48 (Oct. 20) \\
\Ref Narayan, R. 1996a, in Physics of Accretion Disks, eds. S. Kato, 
S. Inagaki, S. Mineshige, J. Fukue (Gordon and Breach), 15 \\
\Ref Narayan, R. 1996b, ApJ, 462, 136 \\
\Ref Narayan, R. 1997, Proc. IAU Colloq. No 163, Accretion Phenomena \& 
Related Outflows, A.S.P. Conf. Series., ed. Wickramasinghe, D.T., 
Ferrario, L. \& Bicknell, G.V. \\
\Ref Narayan, R., Kato, S., \& Honma, F. 1997, ApJ, in press 
(astro-ph/9607019) \\
\Ref Narayan, R., \& McClintock, J. E. 1997, ApJ, in preparation \\
\Ref Narayan, R., McClintock, J. E., \& Yi, I. 1996, ApJ, 457, 821 (NMY) \\
\Ref Narayan, R., \& Yi, I. 1994, ApJ, 428, L13 \\
\Ref Narayan, R., \& Yi, I. 1995a, ApJ, 444, 231 \\
\Ref Narayan, R., \& Yi, I. 1995b, ApJ, 452, 710 \\
\Ref Narayan, R., Yi, I., \& Mahadevan, R. 1995, Nature, 374, 623 \\
\Ref Novikov, I. D., \& Thorne, K. S. 1973, in Blackholes
ed. C. DeWitt \& B. DeWitt (New York: Gordon \& Breach), 345 \\
\Ref Pacholczyk, A. G. 1970, Radio Astrophysics (San Francisco: Freeman) \\
\Ref Paczy\'nski, B., \& Wiita, P. J. 1980, A\&A, 88, 23 \\
\Ref Petrosian, V. 1981, ApJ, 251, 727 \\
\Ref Phinney, E. S. 1981, in Plasma Astrophysics, ed. T. D. Guyenne
\& G. Levy (ESA SP-161), 337 \\
\Ref Piran, T. 1978, ApJ, 221, 652 \\
\Ref Poutanen, J., \& Svensson, R. 1996, ApJ, in press (astro-ph/9605073) \\
\Ref Rees, M. J., Begelman, M. C., Blandford, R. D., \& Phinney, E. S. 1982,
Nature, 295, 17 \\
\Ref Shahbaz, T., Ringwald, F. A., Bunn, J. C., Naylor, T., Charles, P. A.,
\& Casares, J. 1994 MNRAS, 271, L10 \\
\Ref Shapiro, S. I., Lightman, A. P., \& Eardley, D. M. 1976, ApJ, 204, 187 \\
\Ref Smak, J. 1993, Acta Astron., 43, 101 \\
\Ref Spruit, H., Matsuda, T., Inoue, M., \& Sawada, K. 1987, MNRAS, 
229, 517 \\
\Ref Stepney, S., \& Guilbert, P. W. 1983, MNRAS, 204, 1269 \\
\Ref Sunyaev, R. A., \& Titarchuk, L. G. 1980, A\&A, 86, 121 \\
\Ref Svensson, R. 1982, ApJ, 258, 335 \\
\Ref Takahara, F., \& Tsuruta, S. 1982, Prog. Theor. Phys., 67, 485 \\
\Ref Tanaka, Y. \& Shibazaki, N. 1996, ARAA, 34, in press \\
\Ref van Paradijs, J., \& McClintock, J. E. 1995, in X-ray Binaries, ed.
W. H. G. Lewin, J. van Paradijs, \& E. P. J. van den Heuvel (Cambridge:
Cambridge Univ. Press), 58 \\
\Ref Wagner, R. M., Kreidl, T. J., Howell, S. B., \& Starrfield, S. G.
1992, ApJ, 401, L97 \\
\Ref Wagner, R. M., Starrfield, S. G., Hjellming, R. M., Howell, S. B., \&
Kreidl, T. J. 1994, ApJ, 401, L97 \\
\Ref Wheeler, J. C. 1996, in Relativistic Astrophysics, eds. B. Jones
\& D. Markovic (Cambridge Univ. Press) (astro-ph/9606119) \\
\Ref Yi, I. 1996, ApJ, in press (astro-ph/9609146) \\
 
}

\newpage
\begin{appendix}
\section{Non-Spherical Stucture of the ADAF}

Using the results of Narayan \& Yi (1995a) as a guide, we assume that
at each spherical radius $R$, the angular velocity $\Omega$ and the
isothermal sound speed $c_s=(p/\rho)^{1/2}$ are independent of the
polar angle $\theta$.  Consider the condition for ``hydrostatic
equilibrium'' in the $\theta$ direction.  Since the gravitational
force vanishes along this direction, we need consider only the
pressure gradient and the centrifugal term.  This gives
$$
{1\over \rho}{dp\over Rd\theta}=\Omega^2R\sin\theta\cos\theta, \eqno (A.1)
$$
which integrates to
$$
\rho=\rho_0\exp\left[-{\Omega^2R^2\cos^2\theta\over2c_s^2}\right], \eqno (A.2)
$$
We thus obtain the variation of density with polar angle.

Let us assume that the radial velocity $v$ is independent of $\theta$.
(This is not true, but it is an acceptable approximation for the
present purpose).  Then the mass accretion rate corresponding to (A.2)
is
$$
\dot M=4\pi\sqrt{{\pi\over2}}{\rho_0vRc_s\over\Omega}
\;{\rm erf}\left({\Omega R\over\sqrt{2}c_s}\right).\eqno (A.3)
$$
This relation determines $\rho_0$ in terms of $\dot M$.

The global solutions calculated by Narayan, Kato \& Honma (1997) and
Chen, Abramowicz \& Lasota (1997) give $\Omega$, $v$ and $c_s^2$ as
functions of radius $R$.  By substituting these in the above formulae,
we see that we can calculate the density $\rho(R,\theta)$ in the
two-dimensional $(R,\,\theta)$ plane.  Narayan (1997) shows isodensity
contours of a typical solution with $\alpha=0.3$ and $\beta=0.5$.

For the spectral calculations of this paper, a density that varies
with $\theta$ is inconvenient since the various emission processes vary
as different powers of the density.  Therefore, we simplify matters by
replacing the above density profile by one in which the density is
equal to $\rho_0$ over a certain range of angle on either side of the
equator and vanishes near the two poles.  Define the angle $\theta_H$ by
$$
\cos\theta_H=\sqrt{{\pi\over2}}{c_s\over\Omega R}
\;{\rm erf}\left({\Omega R\over\sqrt{2}c_s}\right).\eqno (A.4)
$$
Substituting in eq. (A.3), we see that
$$
\dot M=4\pi\rho_0vR^2\cos\theta_H.\eqno (A.5)
$$

Thus, the density profile (A.2) may be replaced by a simpler model in
which the density is constant over the range $\theta_H<\theta
<\pi-\theta_H$ and vanishes over the poles.  We make use of this
simplified model for the spectral calculations described in the paper.

\end{appendix}

\newpage
\noindent {\large \bf Figure Captions}

\noindent
{Figure 1:} Combined unfolded GIS and SIS spectra of V404 Cyg for the
power law model.

\noindent
{Figure 2:} (a) Allowed grid of variations of the column density
(N$_{\rm H}$) and the power law photon index ($\alpha$) in V404 Cyg.
The contours encompass the 68\%, 95\% and 99\% confidence levels.  (b)
Allowed grid of variations of the normalization flux at 1 keV and
$\alpha$.

\noindent
{Figure 3:} (a) Spectra corresponding to Models 1 and 2 of V404 Cyg in
quiescence (see Tables 2 and 3).  The models correspond to $\beta=0.5$
and 0.95 respectively, with standard values for the other parameters.
The three dots represent measured optical fluxes (from NMY) and the
the error box in the X-ray band, which corresponds to a $2\sigma$
deviation, is obtained from our analysis of ASCA data (\S3.1).  The
mass accretion rates in the two models have been adjusted so that the
calculated spectra pass through the ``throat'' of the error box.  The
peak on the left in the two spectra is due to synchrotron radiation
from the ADAF.  The next bump arises from single Compton scattering,
and the peak on the right is the result of bremsstrahlung emission.
The radiation in the ASCA band is primarily from doubly and triply
Compton scattered photons combined with the low-energy tail of the
bremsstrahlung peak.  (b) Attempts to fit the data with two pure thin
disk models (see text for details).  The models have been adjusted to
pass through the optical data.  The fit in the X-ray band is extremely
poor.

\noindent
{Figure 4:} (a) Spectra corresponding to three models of V404 Cyg in
which the mass of the black hole is varied: $M=8, ~12, ~16 M_\odot$
(Models 3, 1, 4 in Tables 2, 3).  Note that all three models fit the
data very well.  (b) Four models which differ in the transition
radius: $\log r_{\rm tr}=$ 4.9, 4.4, 3.9, 3.4 (Models 5, 1, 6, 7 in Tables
2, 3).  All four models fit the observations well.  However, Model 7
has a large effective temperatures in the disk and may be ruled out
(see \S4).

\noindent
{Figure 5:} (a) Spectra of five models of V404 Cyg which differ in the
choice of viscosity parameter: $\alpha=$ 0.1, 0.2, 0.3, 0.4, 0.5
(Models 8, 9, 1, 10, 11 in Tables 2, 3).  All five models fit the data
very well.  (b) Four models which differ in the assumed direct heating
of the electrons: $\delta = $ 0.001, 0.00316, 0.01, 0.0316 (Models 1,
12, 13, 14 in Tables 2, 3).  All models again fit the data
satisfactorily.  However, the model spectra differ from each other in
the ASCA band, and could in principle be distinguished with more
sensitive observations.

\noindent
{Figure 6:} (a) Spectra of Models 1 and 2 of A0620--00 in quiescence
(see Tables 4, 5).  The models correspond to $\beta=0.5$ and 0.95
respectively, with standard values for the other parameters.  The
optical/UV fluxes and EUV limit are taken from NMY.  The X-ray error
box corresponds to $2\sigma$ limits with fixed N$_{\rm H}$ (see \S3.4 for
details).  We consider both models to be satisfactory.  (b) Three
models of A0620--00 in which the black hole mass is varied:
$M=4.4M_\odot, ~6.1M_\odot, ~12M_\odot$ (Models 3, 1, 4 in Tables 4,
5).

\noindent
{Figure 7:} (a) Three models of A0620--00 in which the transition
radius is varied: $\log(r_{\rm tr})=3.8, ~3.4, ~3.0$ (Models 1, 5, 6
in Tables 4, 5).  Models 5 and 6 have large effective temperatures in
the outer disk and can probably be ruled out (see \S4).  (b) Four
models of A0620--00 which differ in the assumed heating of the
electrons: $\delta=0.001, ~0.00316, ~0.01, ~0.0316$ (Models 1, 7, 8, 9
in Tables 4, 5).  As in the case of V404 Cyg (Fig. 5b), the models
predict different spectral slopes in the X-ray band and could in
principle be distinguished with sensitive observations.

\newpage
\newpage
\centerline{\psfig{figure=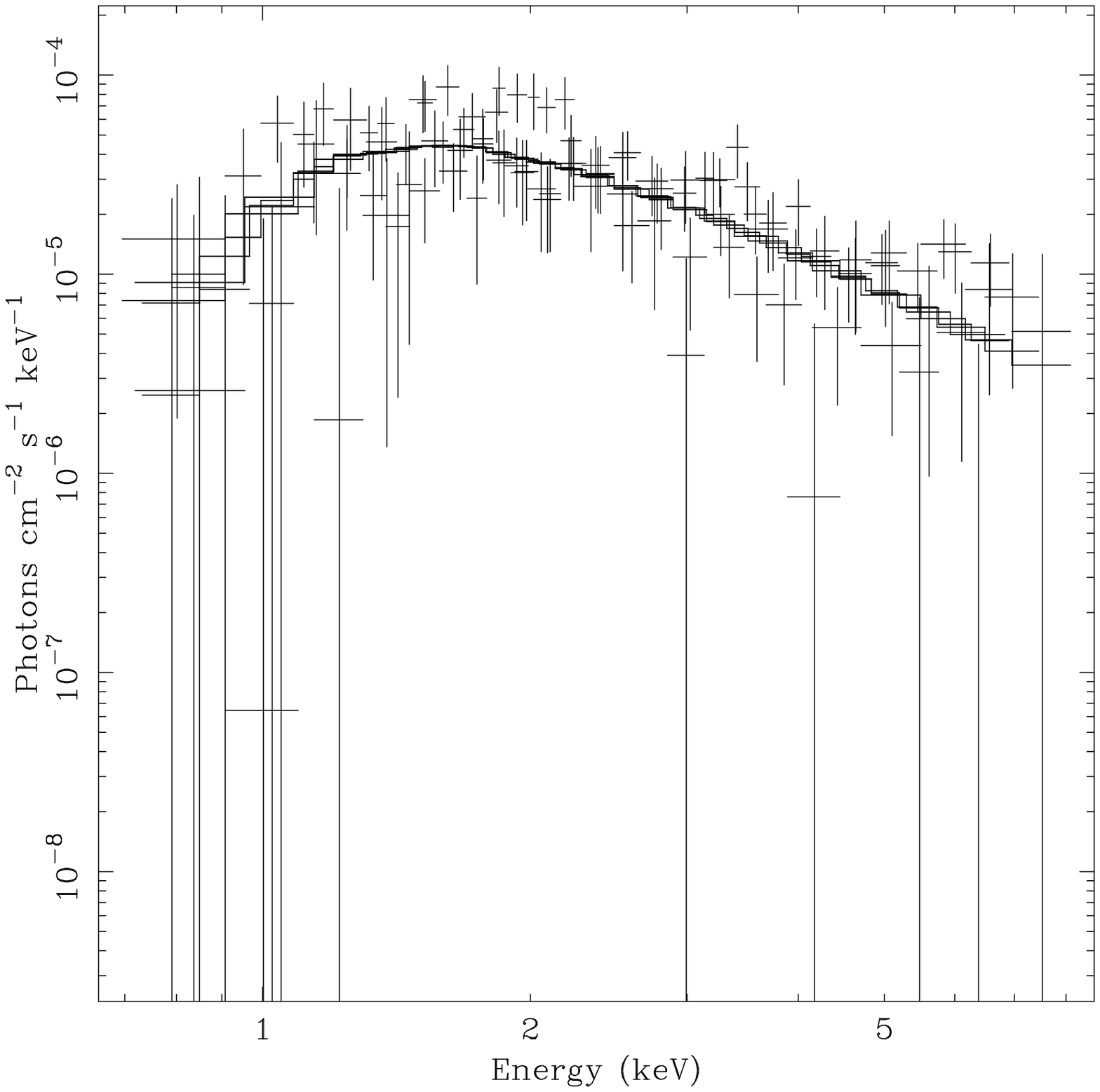,height=13cm,width=12cm}}
\centerline{Figure 1}

\newpage
\psfig{figure=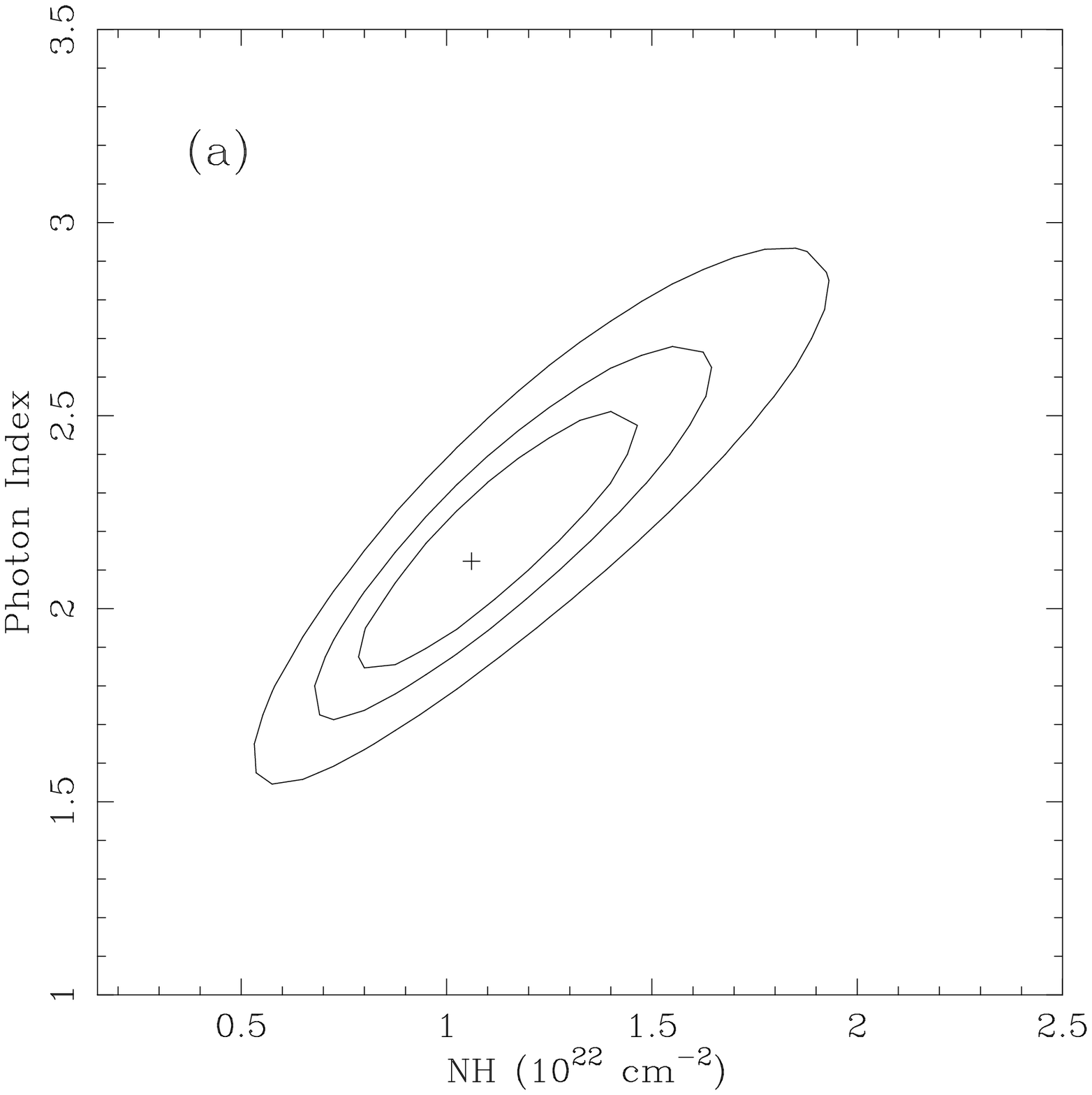,height=15cm,width=15cm}
\centerline{Figure 2a}

\newpage
\psfig{figure=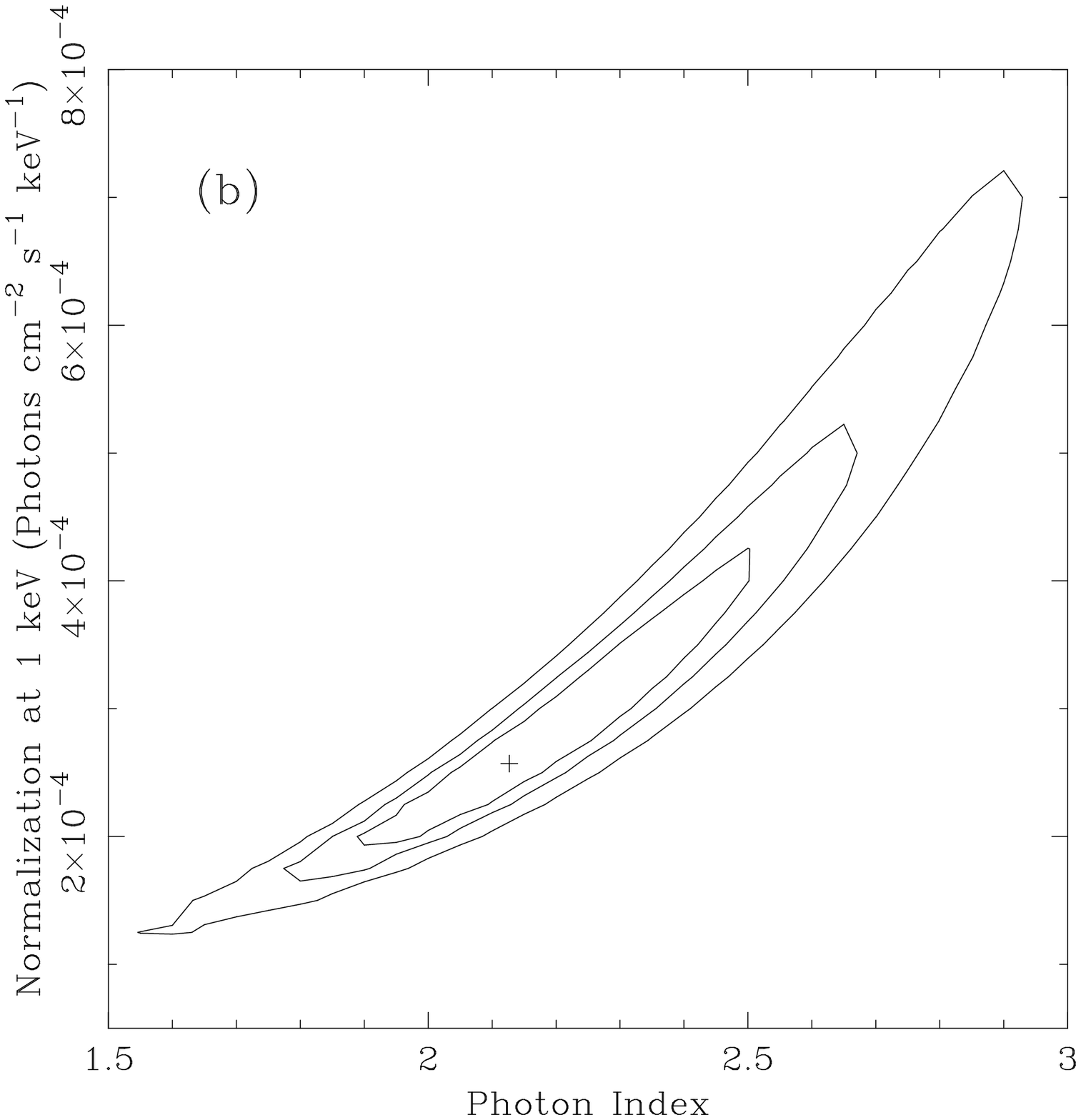,height=15cm,width=15cm}
\centerline{Figure 2b}

\newpage
\centerline{\psfig{figure=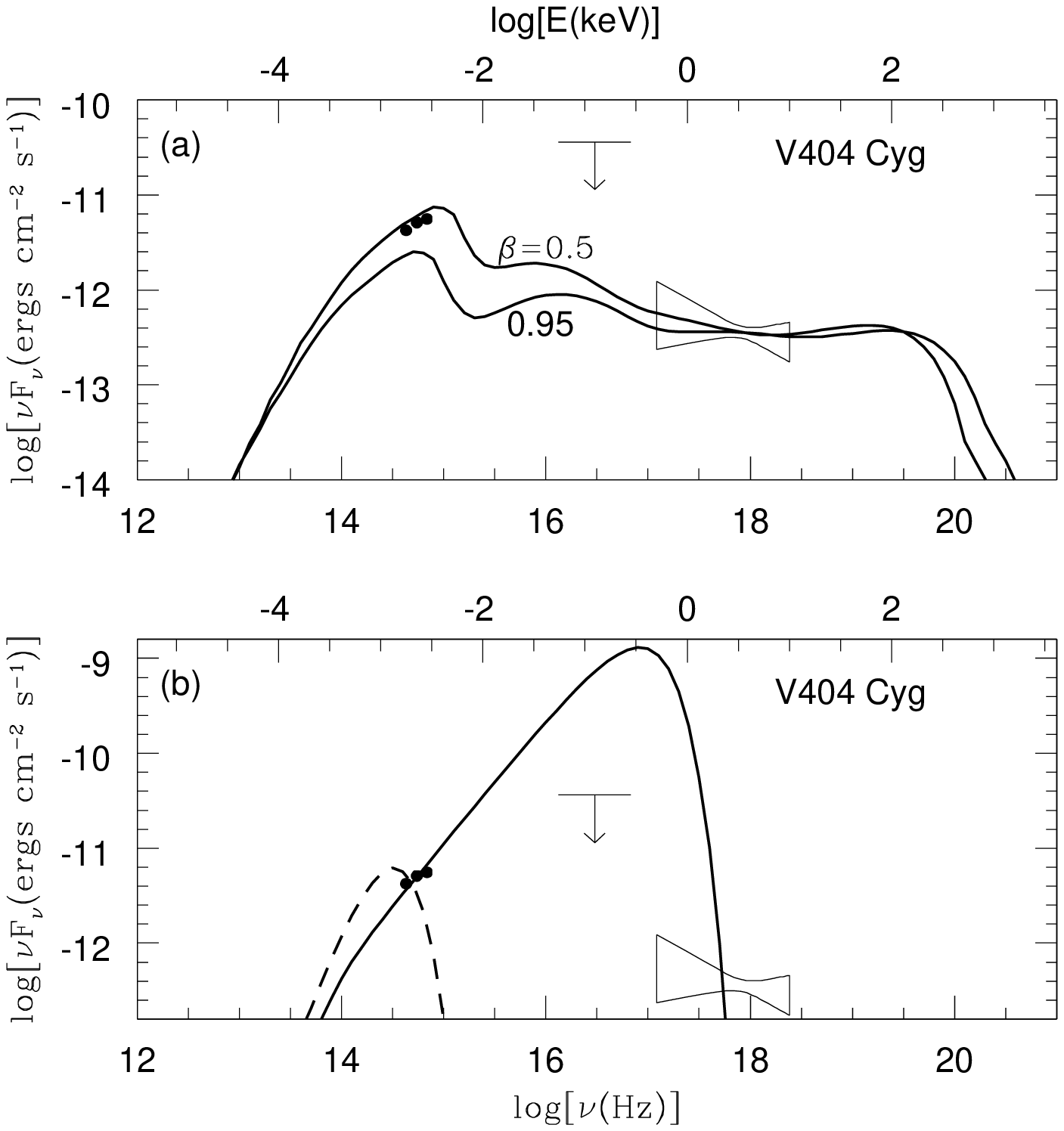,height=20cm,width=20cm}}
\centerline{Figure 3}

\newpage
\centerline{\psfig{figure=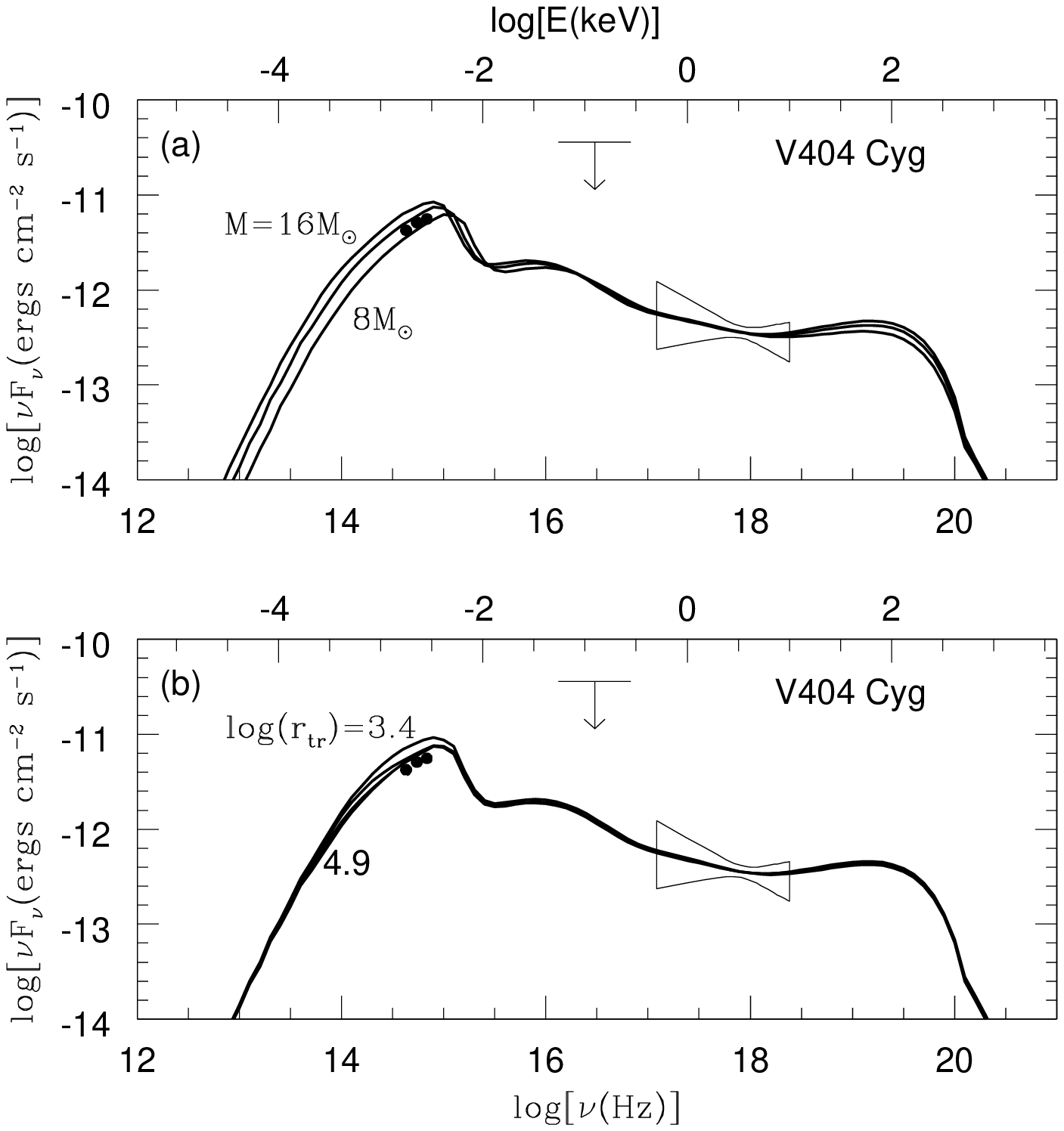,height=20cm,width=20cm}}
\centerline{Figure 4}

\newpage
\centerline{\psfig{figure=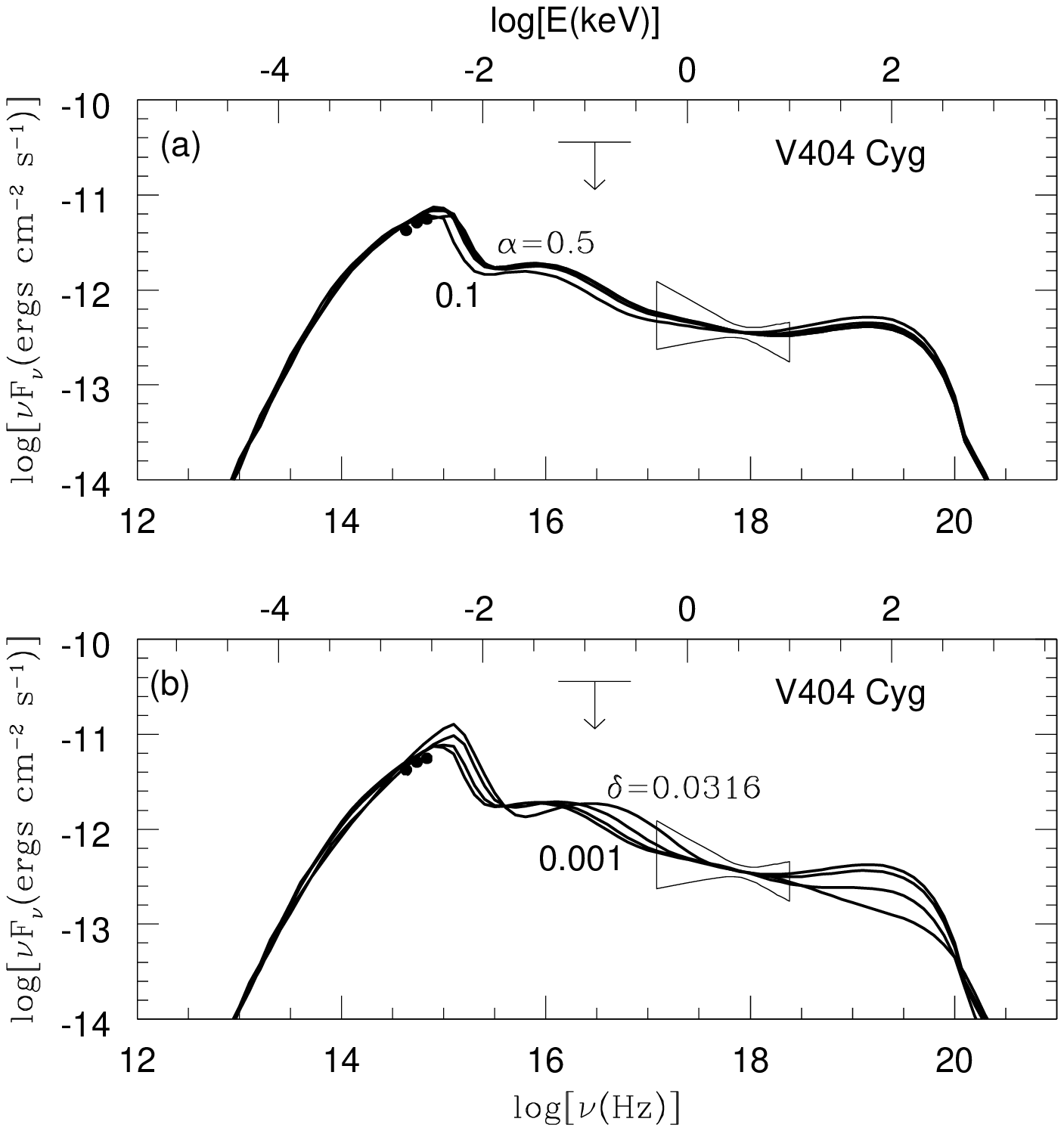,height=20cm,width=20cm}}
\centerline{Figure 5}

\newpage
\centerline{\psfig{figure=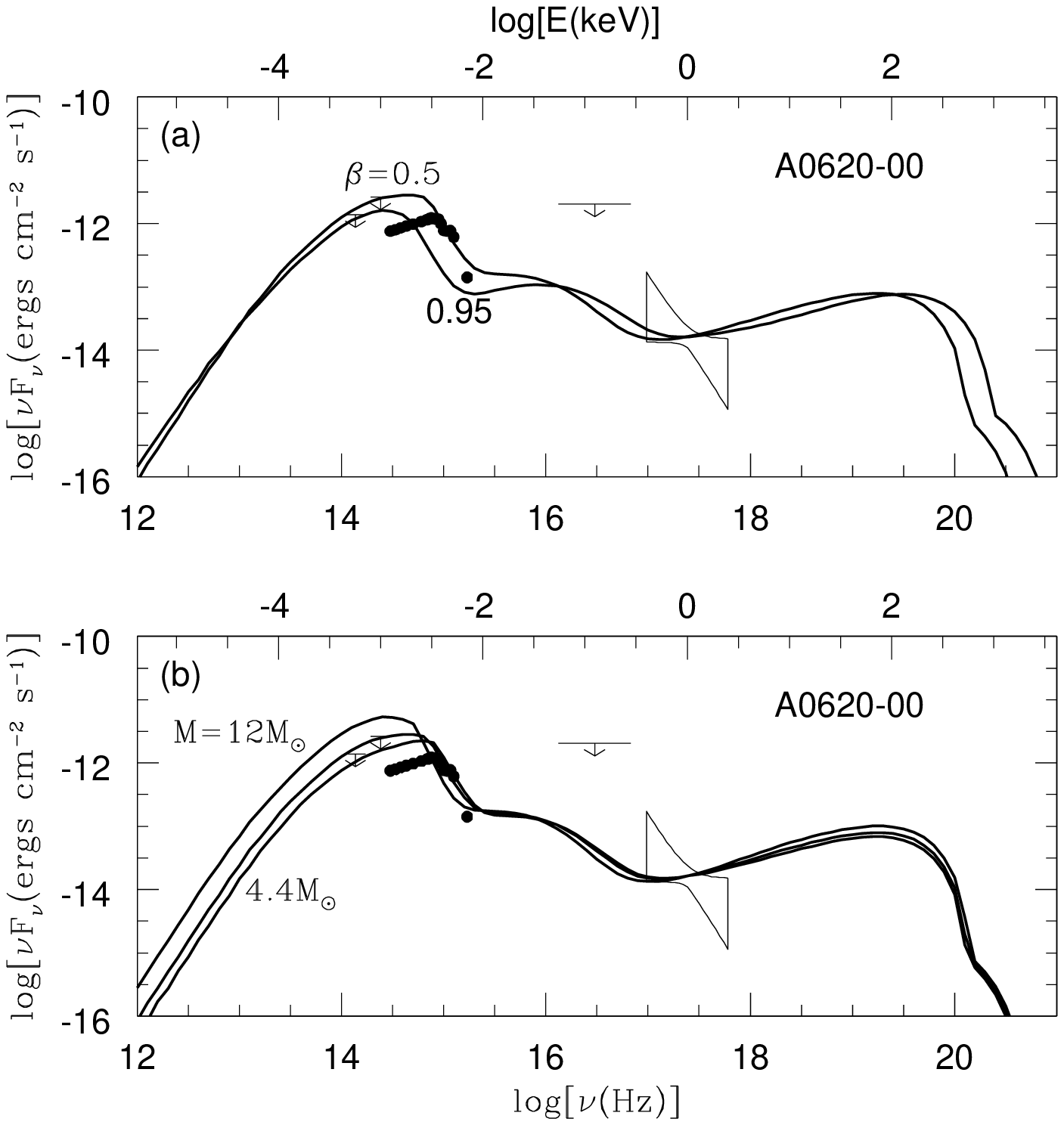,height=20cm,width=20cm}}
\centerline{Figure 6}

\newpage
\centerline{\psfig{figure=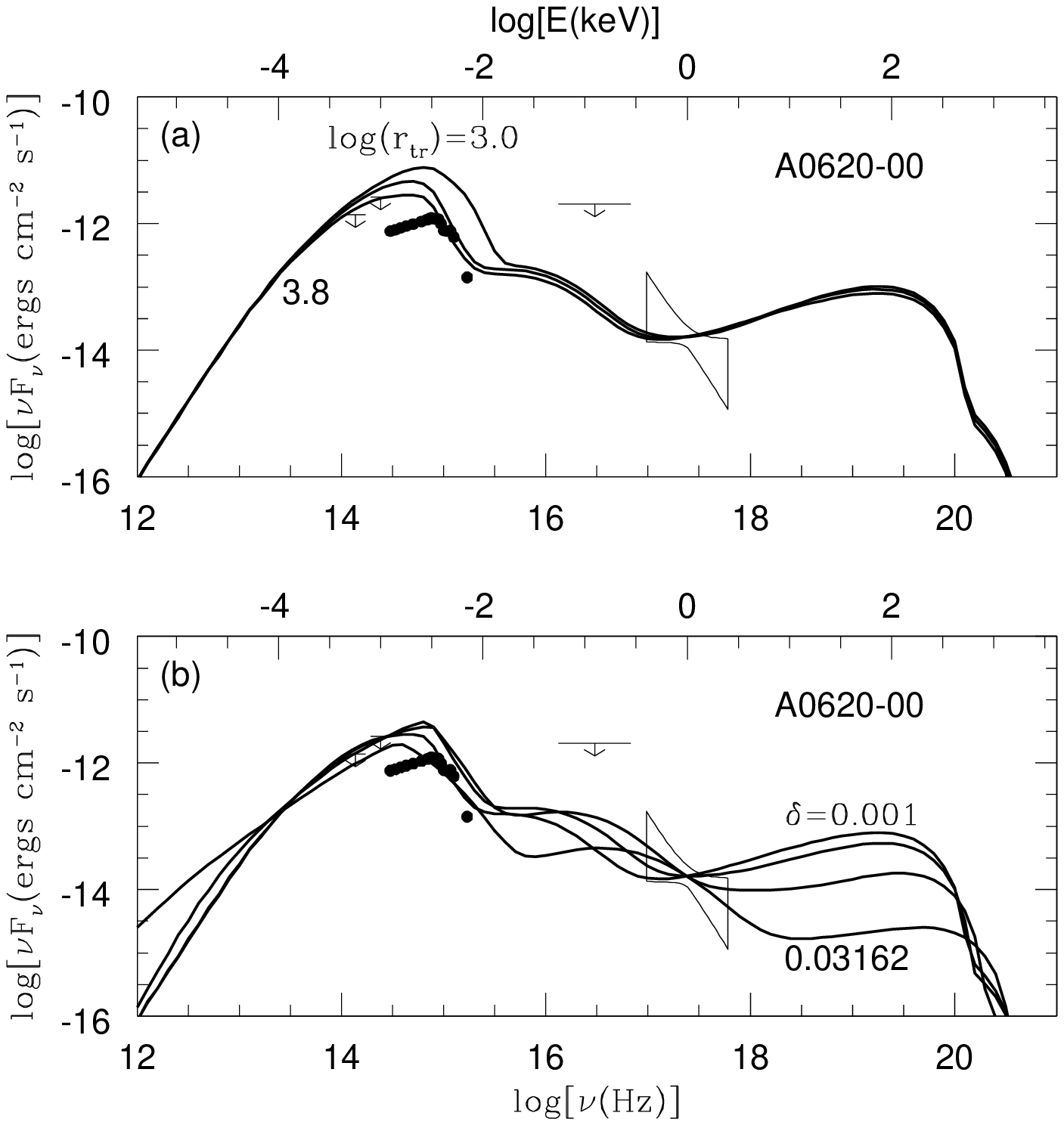,height=20cm,width=20cm}}
\centerline{Figure 7}

\newpage
\begin{table*}[h]
\begin{center}
\begin{tabular}{|ccccl|}
\hline
Model & N$_{\rm H}$ ($\times 10^{22}$ \nh)& $\alpha$, kT & 
Flux\expa @ 1 keV & Red. $\chi^2$ (d.o.f) \\ 
\hline 
PL & $1.10^{+0.30}_{-0.40}$ & $2.10^{+0.50}_{-0.30}$ &
$2.50^{+2.00}_{-0.80}$ & 0.99 (105) \\ 
\hline
TB & $0.81^{+0.30}_{-0.40}$ & $4.60^{+3.60}_{-1.54}$ &
$2.12^{+0.77}_{-0.46}$ & 1.01 (105) \\ 
\hline
BB & $0.18^{+0.22}_{-0.18}$ & $0.82^{+0.08}_{-0.08}$ &
$0.59^{+0.06}_{-0.06}$ & 1.14 (105) \\ 
\hline 
\hline
\end{tabular}
\noindent
\caption{Spectral parameters of V404 Cyg as derived from the archival 40 ksec 
  observation. The fit is performed between 0.7 and 8.5 keV, and the
  uncertainties are computed at the 90\% confidence level for joint
  variation of all parameters.  Models:
  PL: Power Law, TB=Thermal Bremsstrahlung, BB = Blackbody. $\alpha$
  is the photon index. kT is the temperature of either the TB or BB model.}
\end{center}
\expa~$\times 10^{-4}$ \Fkev \\ 
\end{table*}

\begin{table*}[t]
\begin{center}
\begin{tabular}{|cccccccc|}
\hline
Model & M($M_\odot$) & i($^o$) & log($r_{\rm tr}$) & $\alpha$ & $\beta$ & 
$\delta$ & $\dot M$($M_\odot\,$yr$^{-1}$) \\
\hline 
1 & 12 & 56 & 4.4 & 0.3 & 0.5 & 0.001 & $1.22\times10^{-9}$ \\
2 & 12 & 56 & 4.4 & 0.3 & 0.95 & 0.001 & $1.00\times10^{-9}$ \\
\hline
3 & 8 & 56 & 4.4 & 0.3 & 0.5 & 0.001 & $9.12\times10^{-10}$ \\
4 & 16 & 56 & 4.4 & 0.3 & 0.5 & 0.001 & $1.49\times10^{-9}$ \\
\hline
5 & 12 & 56 & 4.9 & 0.3 & 0.5 & 0.001 & $1.22\times10^{-9}$ \\
6 & 12 & 56 & 3.9 & 0.3 & 0.5 & 0.001 & $1.23\times10^{-9}$ \\
7 & 12 & 56 & 3.4 & 0.3 & 0.5 & 0.001 & $1.26\times10^{-9}$ \\
\hline
8 & 12 & 56 & 4.4 & 0.1 & 0.5 & 0.001 & $5.29\times10^{-10}$ \\
9 & 12 & 56 & 4.4 & 0.2 & 0.5 & 0.001 & $8.99\times10^{-10}$ \\
10 & 12 & 56 & 4.4 & 0.4 & 0.5 & 0.001 & $1.52\times10^{-9}$ \\
11 & 12 & 56 & 4.4 & 0.5 & 0.5 & 0.001 & $1.84\times10^{-9}$ \\
\hline
12 & 12 & 56 & 4.4 & 0.3 & 0.5 & 0.00316 & $1.11\times10^{-9}$ \\
13 & 12 & 56 & 4.4 & 0.3 & 0.5 & 0.01 & $8.16\times10^{-10}$ \\
14 & 12 & 56 & 4.4 & 0.3 & 0.5 & 0.0316 & $4.06\times10^{-10}$ \\
\hline
\hline
\end{tabular}
\noindent
\caption{Parameters of models of V404 Cyg shown in Figures 3--5.  Model 1 
is the baseline model, which incorporates the standard values of the
parameters (see \S3.2)}
\end{center}
\end{table*}

\newpage

\begin{table*}[h]
\begin{center}
\begin{tabular}{|cccccc|}
\hline
Model & $L_{\rm ADAF}$(${\rm ergs\,s^{-1}}$) & 
$T_{\rm e,ADAF,max}$(K) & $L_{\rm disk}$(${\rm ergs\,s^{-1}}$) &
$T_{\rm eff,disk,max}$(K) & ${L_{\rm ADAF}+L_{\rm disk}\over\dot Mc^2}$ \\
\hline 
1 & $3.30\times10^{34}$ & $2.83\times10^9$ & $2.52\times10^{32}$ & 1920 & 
$4.81\times10^{-4}$ \\
2 & $1.48\times10^{34}$ & $4.84\times10^9$ & $1.75\times10^{32}$ & 1680 & 
$2.63\times10^{-4}$ \\
\hline
3 & $2.82\times10^{34}$ & $2.77\times10^9$ & $1.99\times10^{32}$ & 2270 & 
$5.49\times10^{-4}$ \\
4 & $3.65\times10^{34}$ & $2.88\times10^9$ & $2.98\times10^{32}$ & 1710 & 
$4.37\times10^{-4}$ \\
\hline
5 & $3.31\times10^{34}$ & $2.83\times10^9$ & $1.80\times10^{31}$ & 1020 & 
$4.80\times10^{-4}$ \\
6 & $3.35\times10^{34}$ & $2.82\times10^9$ & $1.26\times10^{33}$ & 4090 & 
$4.99\times10^{-4}$ \\
7 & $3.48\times10^{34}$ & $2.82\times10^9$ & $4.80\times10^{33}$ & 9490 & 
$5.55\times10^{-4}$ \\
\hline
8 & $2.90\times10^{34}$ & $2.85\times10^9$ & $1.86\times10^{32}$ & 2050 & 
$9.72\times10^{-4}$ \\
9 & $3.21\times10^{34}$ & $2.82\times10^9$ & $2.22\times10^{32}$ & 1930 & 
$6.34\times10^{-4}$ \\
10 & $3.26\times10^{34}$ & $2.83\times10^9$ & $2.88\times10^{32}$ & 1900 & 
$3.83\times10^{-4}$ \\
11 & $3.11\times10^{34}$ & $2.85\times10^9$ & $3.29\times10^{32}$ & 1950 & 
$3.01\times10^{-4}$ \\
\hline
12 & $3.39\times10^{34}$ & $2.89\times10^9$ & $2.26\times10^{32}$ & 1860 & 
$5.44\times10^{-4}$ \\
13 & $3.64\times10^{34}$ & $3.75\times10^9$ & $1.60\times10^{32}$ & 1670 & 
$7.90\times10^{-4}$ \\
14 & $4.19\times10^{34}$ & $5.63\times10^9$ & $7.62\times10^{31}$ & 1380 & 
$1.83\times10^{-3}$ \\
\hline
\hline
\end{tabular}
\noindent
\caption{Luminosities and temperatures of the ADAF and thin disk,
corresponding to the models of V404 Cyg listed in Table 2.  $L_{\rm
ADAF}$ is the total luminosity escaping from the ADAF and corona,
integrated over all directions.  $T_{\rm e,ADAF,max}$ is the maximum
electron temperature in the ADAF.  $L_{\rm disk}$ is the luminosity
escaping from the outer disk.  $T_{\rm eff,max}$ is the maximum
effective temperature in the disk.  The last column gives the overall
efficiency of the accretion flow.}
\end{center}
\end{table*}

\begin{table*}[t]
\begin{center}
\begin{tabular}{|cccccccc|}
\hline
Model & M($M_\odot$) & i($^o$) & log($r_{\rm tr}$) & $\alpha$ & $\beta$ & 
$\delta$ & $\dot M$($M_\odot\,$yr$^{-1}$) \\
\hline 
1 & 6.1 & 55 & 3.8 & 0.3 & 0.5 & 0.001 & $1.31\times10^{-10}$ \\
2 & 6.1 & 55 & 3.8 & 0.3 & 0.95 & 0.001 & $1.20\times10^{-10}$ \\
\hline
3 & 4.4 & 70 & 4.0 & 0.3 & 0.5 & 0.001 & $1.45\times10^{-10}$ \\
4 & 12 & 40 & 3.6 & 0.3 & 0.5 & 0.001 & $1.04\times10^{-10}$ \\
\hline
5 & 6.1 & 55 & 3.4 & 0.3 & 0.5 & 0.001 & $1.41\times10^{-10}$ \\
6 & 6.1 & 55 & 3.0 & 0.3 & 0.5 & 0.001 & $1.49\times10^{-10}$ \\
\hline
7 & 6.1 & 55 & 3.8 & 0.3 & 0.5 & 0.00316 & $1.03\times10^{-10}$ \\
8 & 6.1 & 55 & 3.8 & 0.3 & 0.5 & 0.01 & $4.57\times10^{-11}$ \\
9 & 6.1 & 55 & 3.8 & 0.3 & 0.5 & 0.0316 & $7.46\times10^{-12}$ \\
\hline
\hline
\end{tabular}
\noindent
\caption{Parameters of models of A0620--00 shown in Figures 6 and 7.}
\end{center}
\end{table*}

\newpage

\begin{table*}[h]
\begin{center}
\begin{tabular}{|cccccc|}
\hline
Model & $L_{\rm ADAF}$(${\rm ergs\,s^{-1}}$) & 
$T_{\rm e,ADAF,max}$(K) & $L_{\rm disk}$(${\rm ergs\,s^{-1}}$) &
$T_{\rm eff,disk,max}$(K) & ${L_{\rm ADAF}+L_{\rm disk}\over\dot Mc^2}$ \\
\hline 
1 & $1.04\times10^{33}$ & $2.78\times10^9$ & $1.54\times10^{32}$ & 3710 & 
$1.60\times10^{-4}$ \\
2 & $5.06\times10^{32}$ & $5.04\times10^9$ & $1.40\times10^{32}$ & 3620 & 
$9.51\times10^{-5}$ \\
\hline
3 & $8.97\times10^{32}$ & $2.69\times10^9$ & $6.51\times10^{31}$ & 2930 & 
$1.17\times10^{-4}$ \\
4 & $1.36\times10^{33}$ & $2.93\times10^9$ & $4.22\times10^{32}$ & 4160 & 
$3.03\times10^{-4}$ \\
\hline
5 & $1.17\times10^{33}$ & $2.75\times10^9$ & $4.98\times10^{32}$ & 7520 & 
$2.08\times10^{-4}$ \\
6 & $1.27\times10^{33}$ & $2.73\times10^9$ & $1.42\times10^{33}$ & 15200 & 
$3.18\times10^{-4}$ \\
\hline
7 & $1.33\times10^{33}$ & $3.55\times10^9$ & $1.21\times10^{32}$ & 3490 & 
$2.49\times10^{-4}$ \\
8 & $1.46\times10^{33}$ & $5.65\times10^9$ & $5.37\times10^{31}$ & 2850 & 
$5.86\times10^{-4}$ \\
9 & $7.40\times10^{32}$ & $1.24\times10^{10}$ & $9.11\times10^{30}$ & 1840 & 
$1.77\times10^{-3}$ \\
\hline
\hline
\end{tabular}
\noindent
\caption{Luminosities and temperatures of the ADAF and thin disk,
corresponding to the models of A0620--00 listed in Table 4.  $L_{\rm
ADAF}$ is the total luminosity escaping from the ADAF and corona,
integrated over all directions.  $T_{\rm e,ADAF,max}$ is the maximum
electron temperature in the ADAF.  $L_{\rm disk}$ is the luminosity
escaping from the outer disk.  $T_{\rm eff,max}$ is the maximum
effective temperature in the disk.  The last column gives the overall
efficiency of the accretion flow.}
\end{center}
\end{table*}

\end{document}